\documentclass[sigconf]{aamas} 

\usepackage{amsmath, amsfonts}
\usepackage{array}
\usepackage{balance} 

\setcopyright{ifaamas}
\acmConference[AAMAS '23]{Proc.\@ of the 22nd International Conference
on Autonomous Agents and Multiagent Systems (AAMAS 2023)}{May 29 -- June 2, 2023}
{London, United Kingdom}{A.~Ricci, W.~Yeoh, N.~Agmon, B.~An (eds.)}
\copyrightyear{2023}
\acmYear{2023}
\acmDOI{}
\acmPrice{}
\acmISBN{}

\acmSubmissionID{1014}

\title{Data Structures for Deviation Payoffs}

\author{Bryce Wiedenbeck}
\affiliation{\institution{Davidson College} \country{United States}}
\email{brwiedenbeck@davidson.edu}

\author{Erik Brinkman}
\affiliation{\institution{Independent Researcher} \country{United States}}
\email{erik.brinkman@gmail.com}

\begin{abstract}
We present new data structures for representing symmetric normal-form games.
These data structures are optimized for efficiently computing the expected utility of each unilateral pure-strategy deviation from a symmetric mixed-strategy profile.
The cumulative effect of numerous incremental innovations is a dramatic speedup in the computation of symmetric mixed-strategy Nash equilibria, making it practical to represent and solve games with dozens to hundreds of players.
These data structures naturally extend to role-symmetric and action-graph games with similar benefits.
\end{abstract}


\newcommand{\PreserveBackslash}[1]{\let\temp=\\#1\let\\=\temp}
\newcolumntype{C}[1]{>{\PreserveBackslash\centering}p{#1}}
\newcolumntype{R}[1]{>{\PreserveBackslash\raggedleft}p{#1}}
\newcolumntype{L}[1]{>{\PreserveBackslash\raggedright}p{#1}}

\newcommand{\prof}{\vec{s}}
\newcommand{\config}{\vec{c}}
\newcommand{\reps}[1]{\textrm{Reps} \left( #1 \right)}
\newcommand{\mix}{\vec{\sigma}}
\newcommand{\purepay}{v}
\newcommand{\devpay}{\vec{u}}
\newcommand{\reg}{\textrm{reg}}
\newcommand{\devderiv}[3]{\frac{\partial \devpay_{#2}(#1)}{\partial #3}}
\newcommand{\logrepdevpay}{\lambda}
\newcommand{\devcont}{\gamma}

\newcommand{\bigdot}{\mbox{\LARGE.}}

\newcommand{\threevec}[3]{\begin{bmatrix} #1 \\ #2 \\ #3 \end{bmatrix}}
\newcommand{\angles}[1]{\left< #1 \right>}

\begin{document}


\pagestyle{fancy}
\fancyhead{}


\maketitle 


\section{Introduction}

Symmetric games, where all players face identical incentives, play a central role in game-theoretic analysis.
Many of the classic examples used to teach game-theoretic concepts \cite{essentials_of_game_theory, multiagent_systems} such as prisoner's dilemma and rock-paper-scissors are symmetric, and the original paper proposing the concept of a Nash equilibrium \cite{Nash} addressed symmetry and proved that symmetric games must have symmetric equilibria.
Symmetric games are quite common in the recent multi-agent systems literature \cite{wang2021spoofing, boix2020multiplayer}.
The role of symmetry becomes especially prominent in games with a large number of players, because it is often only with the help of player symmetry that incentives can even be tractably described, and multiple distinct sub-fields of computational game theory \cite{EGTA, MFG} rely heavily on foundational assumptions about player symmetry to scale up game-theoretic analysis.

Despite this importance, data structures for representing and solving large symmetric games have received a paucity of attention in the research literature.
Most libraries for representing and solving games, particularly Gambit \cite{Gambit}, but also GameTracer \cite{GameTracer} and QuantEcon \cite{QuantEcon}, include few tools for efficiently solving large symmetric games.
One notable exception is the work of \citet{action-graph_games} on action-graph games, which incorporate player symmetry, and which have been partially implemented in Gambit.
However, action-graph games focus primarily on theoretical compactness in the action space, and only consider expected utility calculations enough to ensure they take polynomial time.
The data structures that have been implemented are not specifically optimized for equilibrium computation under player symmetry, meaning that solving large symmetric games remains inefficient.

In the sub-fields that scale up symmetric game analysis, approximations play a central role.
Simulation-based games \cite{EGTA} frequently employ player reduction methods \cite{DPR} that use a reduced game with a very small number of players to replace a game with dozens or hundreds of players in the hope that analysis of the reduced-game will approximately hold in the large game.
Mean-field games \cite{MFG} give up on representing discrete players and replace them with an average effect that aggregates over a continuum.
Other approaches aim to identify underlying representational compactness \cite{twins_reduction, structure_learning} or to avoid explicitly constructing game models in favor of sampling-based techniques \cite{ADIDAS, DPLearn}.
These approximations are sometimes unavoidable, when the representation is too large or the payoff computations are too slow for a game to even be enumerated, but analysts often end up resorting to approximation when faced with any non-trivial number of players.
The result is a substantial gap between the very small games that can be represented and solved exactly and the very large games where approximation is necessary.
We bridge this gap by designing efficient data structures that make it practical to exactly solve much larger instances.

In this work, we present a detailed exploration of data structure improvements for symmetric normal-form games, with specific focus on the task of computing symmetric mixed-strategy Nash equilibria.
We argue, following \citet{notes_on_symmetric_equilibria}, that this is by far the most compelling solution concept for symmetric games, because symmetric equilibria are guaranteed to exist, and it provides greater intuitive explanatory power for symmetries of the game to be reflected in its equilibria.
While computing symmetric Nash equilibria can be hard in the worst case \cite{Daskalakis_nash_complexity, Conitzer_nash_complexity}, incomplete local-search algorithms such as replicator dynamics and gradient descent are often highly successful in practice.

To facilitate algorithms for identifying symmetric equilibria, we aim for data structures that optimize the calculation of \emph{deviation payoffs} (and their derivatives).
For a symmetric mixed strategy employed by all players, the deviation payoff vector gives, for each action, the expected utility a single agent would receive if they deviated unilaterally to that action.
For a large fraction of the algorithms used to compute Nash equilibria, the ability to evaluate deviation payoffs for symmetric mixed strategies is both necessary and sufficient; for most other algorithms, deviation payoffs plus their derivatives suffice.
We show that focusing the design of symmetric game data structures toward deviation payoffs leads to a number of optimizations that jointly yield a dramatic improvement in the practical efficiency of solving large symmetric games.

\subsection{Contributions}

We describe seven distinct upgrades to classic data structures for representing symmetric normal-form games.
Two of these (sections \ref{sec:pre-reps} and \ref{sec:opp-config}) constitute asymptotic improvements:
changing from $P$-player profiles to $(P-1)$-opponent configurations reduces the number of stored payoff entries (with $A$ actions) from $A\binom{P+A-1}{P}$ to $A \binom{P+A-2}{P-1}$,
and pre-computing probability weights to avoid repeated multinomial calculations accelerates equilibrium searches by a factor of $A$.
Four of the upgrades focus on vectorization, resulting in better constants and enabling SIMD acceleration.
And we find---as has been well-established in the settings of neural networks and scientific computation---that such improvements can qualitatively change the scope of the computational tools.
The remaining upgrade (section~\ref{sec:log-probs}) follows because the overall expansion-of-scope enables us to analyze games so large that the probability calculations can overflow 64-bit integers, necessitating a switch to a log-space representation of payoffs and probabilities.

Our main result is a roughly ten-thousand-fold speedup in the running time of equilibrium computation algorithms for symmetric games with many players.
This makes it possible to run practical-but-incomplete search methods like replicator dynamics on any symmetric game that can fit in memory, and can also facilitate other slower equilibrium-search techniques.
Our results effectively close the gap between small and large symmetric games, relegating approximation techniques to only those games too large to be represented explicitly.

Two open-source libraries implement most of the data structures and algorithms we discuss.
The \texttt{gameanalysis.jl}\footnote{\url{https://github.com/Davidson-Game-Theory-Research/gameanalysis.jl}} repository provides simple Julia implementations of the best CPU and GPU versions of our symmetric game data structure, and also includes all of the code for experiments in this paper.
The \texttt{gameanalysis.py}\footnote{\url{https://github.com/egtaonline/gameanalysis}} module implements in Python the data structure variants for role-symmetric games that we discuss in Section~\ref{sec:role-symmetry} and provides a richer command-line interface with numerous tools for solving role-symmetric games.

\section{Background}

\subsection{Terminology}

In a symmetric game, all players have the same action set and identical incentives.
We therefore refer to the number of players $P$, but rarely distinguish individual players.
We call the number of actions $A$, and will often index the actions by $a \in \{ 1, \ldots, A\}$.

A \emph{profile} specifies an action for each player, and in a symmetric game, we can represent a profile by an integer-vector $\prof$ specifying a non-negative number of players selecting each action.
We denote the entries in a vector with a subscript, so $\prof_a$ is the player-count for action $a$.
We use a superscript $i$ or $j$ to index vectors in a collection: for example $\prof^i$; to avoid ambiguity, exponents are generally applied to a quantity in parentheses.
We will also distinguish a profile from an \emph{opponent configuration} $\config$, which differs only in that $\sum_a \prof_a = P$, while $\sum_a \config_a = P-1$.
We refer to the configuration resulting from removing action $a$ from profile $\prof$ as $(\prof \mid a)$, since it will often appear in probability calculations and other contexts where it is \emph{given} that one player selects action $a$.
In terms of the integer-vector representation, $(\prof \mid a)$ subtracts 1 from dimension $a$ of $\prof$.
A symmetric game's payoffs can be expressed as the value achieved by a player choosing action $a$ when opponents play configuration $\config$, which we denote by $\purepay_a \left( \config \right)$ normally, or by $\purepay_a \left( \prof \mid a \right)$, when working in terms of profile $\prof$.

A mixed strategy specifies a probability distribution over actions, and a mixed strategy used by all players is called a symmetric mixed-strategy profile.
We denote this with the variable $\mix$, and will often abbreviate the term by referring to a \emph{mixture}.
When computing probabilities, we will frequently refer to the number of asymmetric arrangements corresponding to a symmetric configuration, which we call \emph{repetitions}.
We denote this quantity by $\reps{\config}$ or $\reps{\prof \mid a}$, and calculate it with the following multinomial:

\begin{equation}
\label{eq:dev_reps}
\reps{\config} = \binom{P-1}{\config_1, \config_2, \ldots, \config_A} = \frac{(P-1)!}{\config_1! \config_2! \ldots \config_A!}
\end{equation}

\subsection{Deviation Payoffs}

When analyzing a symmetric game, we are most often interested in computing symmetric mixed-strategy Nash equilibria.
For many algorithms that compute such equilibria a necessary and sufficient condition is the ability to compute deviation payoffs, and for most other algorithms, deviation payoffs plus deviation derivatives suffice.
We begin by formally defining these terms, and then describe their application to our preferred equilibrium computation algorithms in section~\ref{sec:eq_comp}.

Given a symmetric mixed-strategy profile $\mix$, we define the \emph{deviation payoff} $\devpay_a(\mix)$ for action $a$ as the expected utility one player would receive if they played $a$ while all opponents randomized according to $\mix$.
This expectation is often expressed as a sum over all profiles in which $a$ is played of the probability that profile occurs times the payoff to $a$ in that profile:

\begin{equation}
\label{eq:prof_dev_pays}
\devpay_a(\mix) = \sum_{\prof : \, \prof_a > 0} \Pr_{\mix} \left( \prof \mid a \right) \purepay_a \left( \prof \mid a \right)
\end{equation}

\noindent
but can be stated much more cleanly using configurations:

\begin{align}
\label{eq:config_dev_pays}
\devpay_a(\mix) &= \sum_{\config} \purepay_a \left( \config \right) \Pr_{\mix} \left(\config \right) \\
&= \sum_{\config} \purepay_a \left( \config \right) \reps{ \config } \prod_{a'} \left( \mix_{a'} \right)^{\config_{a'}}
\label{eq:config_reps_dev_pays}
\end{align}

\noindent
The deviation payoff vector $\devpay(\mix)$ collects these values for all actions $a \in \{ 1, \ldots, A\}$.
We call the partial derivatives of deviation payoffs with respect to mixture probabilities \emph{deviation derivatives}.
Specifically, $\frac{\partial \devpay_a(\mix)}{\partial \mix_s}$ gives the change in the deviation payoff for action $a$ as the probability of action $s$ is varied.
Again this can be expressed in terms of profiles, but is more straightforward in terms of configurations:
\footnote{\label{note:div_zero}Note that efficient computation of $\left( \mix_{s}\right)^{\config_s - 1}$ can result in numerical errors for mixtures where $\mix_s = 0$.
This sort of error can be avoided here and elsewhere with no real loss of precision by lower-bounding mixture probabilities at machine-epsilon.}

\begin{equation}
\label{eq:dev_deriv}
\devderiv{\mix}{a}{\mix_s} = \sum_{\config} \purepay_a(\config) \reps{\config} \left( \config_s \right) \left( \mix_{s}\right)^{\config_s - 1} \prod_{a' \ne s} \left( \mix_{a'} \right)^{\config_{a'}}
\end{equation}

We can define most other standard game-theoretic quantities for symmetric games in terms of deviation payoffs.
The expected utility experienced by all players when following a symmetric mixed strategy is given by the dot product $u(\mix) = \devpay(\mix) \cdot \mix$.
The regret of a mixture is $\reg(\mix) = \max_a \left( \devpay_a(\mix) - u(\mix) \right)$.
A symmetric Nash equilibrium is a mixture with $\reg(\mix) = 0$, while an approximate Nash equilibrium has $\reg(\mix) \le \varepsilon$, for suitable $\varepsilon$.

\subsection{Equilibrium Computation}

\label{sec:eq_comp}

\begin{figure}[t]
\center
\includegraphics[width=\columnwidth]{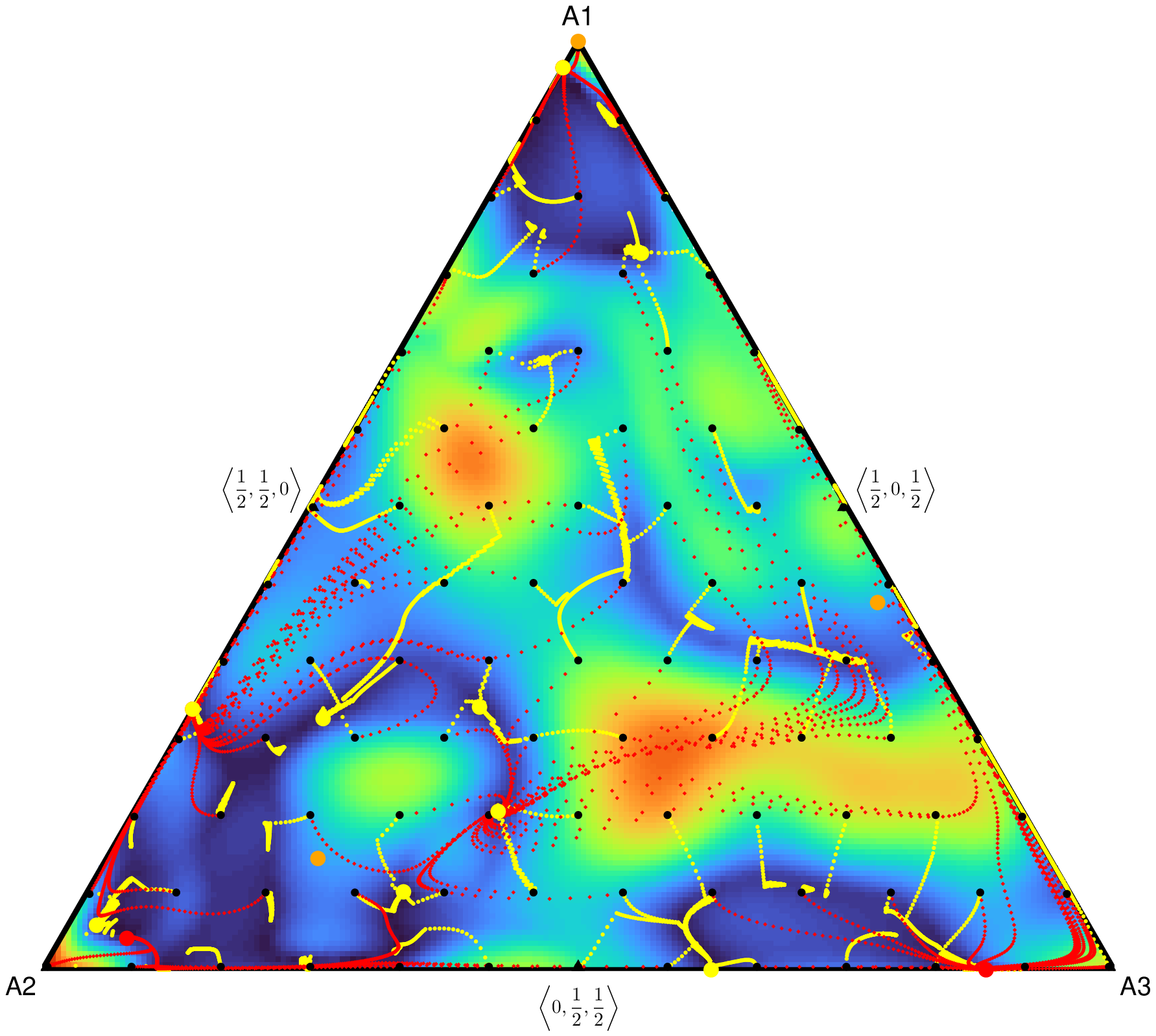}
\caption{Execution traces for both replicator dynamics (red) and gradient descent (yellow) on a 100-player, 3-action game.
The underlying heatmap shows mixture regret.
Black points are starting mixtures.
Large points indicate $\varepsilon$-equilibria.
The 3 orange equilibria were found by both algorithms.
RD found 2 unique equilibria, while GD found 6.}
\label{fig:nash_traces}
\end{figure}

Computing deviation payoffs and/or their derivatives is the key step for a number of algorithms that identify Nash equilibria in symmetric games.
We describe two of the most practical algorithms here: replicator dynamics, which depends on deviation payoffs, and gradient descent on sum-of-gains, which depends on deviation derivatives.
Our data structures can also efficiently support a number of other Nash algorithms, including fictitious play \cite{fictitious_play}, Scarf's simplicial subdivision \cite{Scarf}, and the global Newton method of \citet{global_newton}, as well as some algorithms for correlated equilibria \cite{compact_correlated}.
The Nash Algorithms Appendix\footnote{See Appendix C: Nash Algorithms. \url{https://arxiv.org/abs/2302.13232}} presents further details on how some of these algorithms depend on deviation payoffs.

\paragraph{Replicator dynamics} \cite{replicator_dynamics} is often presented as a rule governing evolutionary dynamics, but can also be viewed as an algorithm for computing symmetric Nash equilibria.
Expressed in terms of deviation payoffs, replicator dynamics starts from some initial mixture $\mix^0$ at $t=0$, and performs iterative updates of the form:

\begin{align*}
\vec{w}_a &\gets \mix_a^{t} \ \devpay_a(\mix^{t}) \\
\mix_a^{t+1} &\gets \frac{\vec{w}_a}{\sum_{a^\prime}\vec{w}_{a^\prime}}
\end{align*}

\noindent
This update assumes that all payoffs are non-negative; a positive affine transformation can be applied to any game to ensure this assumption holds (and to adjust the effective step-size).

\paragraph{Gradient descent} is a classic local-search algorithm for minimizing differentiable functions.
We can easily define a function whose minima correspond to Nash equilibria based on the following sum of deviation gains:

\begin{equation*}
g(\mix) = \sum_a \max \left( 0 , \devpay_a(\mix) - \devpay(\mix) \cdot \mix \right) 
\end{equation*}

\noindent
Then we can iteratively take steps in the direction of $-\nabla g(\mix)$.
The elements of this gradient vector are given by:

\begin{equation*}
\nabla_s(g) = \sum_{a} \left( \devderiv{\mix}{a}{s} - \devpay_s(\mix) - \sum_{a'} \mix_{a'} \devderiv{\mix}{a'}{s} \right) 1_{g_s}
\end{equation*}

\noindent
Where $1_{g_s}$ is an indicator variable for ${\devpay_s(\mix) > \devpay(\mix) \cdot \mix}$.
So deviation payoffs and deviation derivatives suffice to compute the gain gradient.
When performing gradient descent, the mixture resulting from $\mix - \nabla g(\mix)$ may not lie on the probability simplex, so it is necessary to project each step back onto the simplex, which we do using the method from \citet{simplex_projection}.

Neither replicator dynamics nor gradient descent is guaranteed to identify a Nash equilibrium.
However, since these algorithms are many orders of magnitude faster than complete algorithms like simplicial subdivision or global Newton, it is practical to re-run them from many initial mixtures.
We find that both algorithms tend to identify multiple equilibria in large symmetric games and that the sets of equilibria they find are often only partially overlapping.
On the other hand, we find that fictitious play and other best-response-based updates are ineffective on large symmetric games.
Therefore in practice, we recommend repeatedly running both replicator dynamics and gradient descent, filtering by regret to isolate $\varepsilon$-Nash mixtures, and merging the resulting equilibrium sets.
An example trace from running both algorithms on a 100-player, 3-action Gaussian mixture game (represented using our data structures) appears in Figure~\ref{fig:nash_traces}.

\section{Data Structure Improvements}

The classic payoff-matrix representation of a normal-form game has a dimension for each player, and a size along each dimension equal to the number of actions available to that player.
In a symmetric game, it suffices to store just one player's payoffs, so if there are $P$ players and $A$ actions, a symmetric game can be represented by a symmetric tensor of size $A^P$.

\begin{figure}
\includegraphics[width=\columnwidth]{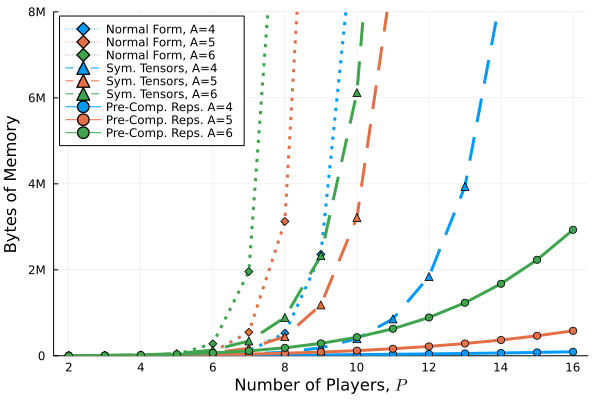}
\caption{The normal-form (dotted lines) and symmetric-tensor (dashed lines) representations use significantly more memory than the largest of our proposed data structures.}
\label{fig:nfg_memory}
\end{figure}

Because symmetric tensors also arise in other settings, it is worth considering whether generic techniques for symmetric tensors would suffice for efficiently representing symmetric games.
In particular, \citet{symmetric_tensors} propose a block-representation of symmetric tensors and a cache-conscious algorithm for multiplying with them.
In principle, their representation has the same level of asymptotic compression as the data structures presented here.
However, in Figure~\ref{fig:nfg_memory}, we compare the memory footprint of the \citeauthor{symmetric_tensors} symmetric-tensor representation (as implemented by the \texttt{SymmetricTensors.jl} library) against the largest of our symmetric-game data structures (pre-computed repetitions), showing that while the block-symmetric representation is better than storing the full payoff matrix, it is significantly larger than any of our data structure variants.

\subsection{Payoff Dictionary}

We therefore turn to purpose-built data structures for storing symmetric games.
The classic approach is to store a mapping from profiles to payoffs.
In Gambit \cite{Gambit} the implementation of symmetric action-graph games stores this mapping in a trie, while others have used a database \cite{EGTAOnline}.
Such a mapping is also implicit in many theoretical descriptions, for example when relating the size of a compact representation to the number of profiles \cite{action-graph_games, resource-graph_games}.
In our experiments, our baseline data structure stores this mapping using a hash table that maps a vector of integers representing the profile $\prof$ to a vector of floats that stores the payoff $\purepay_a(\prof)$ for each played action with $\prof_a > 0$.
With $P$ players and $A$ actions, this table will store $\binom{P+A-1}{P}$ profiles.

Calculating a deviation payoff $u_a(\vec{\sigma})$ using this mapping requires iterating through all profiles to compute the sum from Equation~\eqref{eq:prof_dev_pays}, where the probabilities are given by this expression:

\begin{equation}
\label{eq:prof_prob}
\Pr_{\mix} \left( \prof \mid a \right) = \reps{\prof \mid a} \left( \mix_a \right)^{\prof_a - 1} \prod_{a' \ne a} \left( \mix_{a'} \right)^{\prof_{a'}}
\end{equation}

\noindent
$\reps{\prof \mid a}$, calculated according to the multinomial from Equation~\eqref{eq:dev_reps}, gives the number of asymmetric orderings of the symmetric configuration $\left( \prof \mid a \right)$, while the remaining terms gives the probability of the opponents jointly playing one such asymmetric ordering.
The Worked Examples Appendix shows the full mapping and a detailed walk-through of the deviation payoff calculation on an example 3-player, 3-strategy symmetric game for this version of the data structure as well as for all subsequent variants.
We strongly recommend stepping through these examples to help resolve any confusion that arises from the data structure and algorithm descriptions that follow.\footnote{\label{note:worked_examples}See Appendix A: Worked Examples. \url{https://arxiv.org/abs/2302.13232}}

\subsection{Array Vectorization}

The first idea for improving this deviation payoff calculation is to store the profiles $\prof$ and the payoffs $\purepay$ in a pair of two-dimensional arrays with parallel structure, denoted by a shared index $i$ over all profiles.
In both arrays, each row corresponds to an action and each column corresponds to a profile, resulting in arrays of size $A \times \binom{P+A-1}{P}$.
Extracting column $i$ of the profiles-array gives a profile vector $\prof^i$ with the count for each action.
The corresponding column of the payoffs-array stores the payoff $\purepay_a \left( \prof^i \mid a \right)$ for each action $a$ where that profile has a non-zero count.

Of note, the array representation here (and in all subsequent data structure variants) hinders direct look-up of payoffs for a pure-strategy profile using the mapping.
However, this trade-off is clearly worthwhile for several reasons:

\begin{enumerate}
\item Calculating mixed-strategy deviation payoffs is a much tighter bottleneck than looking up pure-strategy payoffs.
\item Symmetric mixed-strategy equilibria in symmetric games are more often relevant than asymmetric pure-strategy equilibria (and are guaranteed to exist).
\item The profile-to-index mapping can be computed by a ranking algorithm for combinations-with-replacement \cite{Combinatorics}, or can be stored alongside the arrays for fast lookup with only linear additional memory.
\end{enumerate}

Using this array representation, we can vectorize each of the steps in the calculation of the deviation payoff $\devpay_a(\mix)$.
We describe each of the following operations in terms of profile $i$, and broadcast those operations across array columns.
First, we can compute a mask $m$ identifying the profiles where action $a$ is played:
\begin{equation}
\label{eq:mi}
    m^i \gets \prof^{i}_{a} \ne 0
\end{equation}
Then for each such profile, we can remove a player choosing $a$ to get the opponent-configuration $\config^i = (\prof^{i} | a)$ by subtracting an indicator vector $\vec e_{a}$ for action $a$:
\begin{equation}
\label{eq:ci}
    \config^i \gets \prof^i - \vec e_{a}
\end{equation}
For each profile, the probability $p^i = \Pr \left( \config^i \right)$ of the opponent configuration is calculated by a multinomial to compute repetitions along with a broadcast-and-reduce to apply exponents and take the product, resulting in an array of configuration probabilities:
\begin{equation}
\label{eq:pi}
    p^i \gets \reps{\config^i} \prod_{a^\prime} \left( \mix_{a^\prime} \right)^{\config^{i}_{a^\prime}}
\end{equation}
Finally the deviation payoff for action $a$ can be computed by extracting row $a$ of the payoffs array, multiplying element-wise by the configuration probabilities, masking out the invalid configurations, and summing over the remaining profiles:\footnotemark[5]
\begin{equation*}
    \devpay_{a}(\mix) \gets \sum_i m^i\ p^i\  \purepay_{a}\left(\config^i \right)
\end{equation*}

Since these steps can be performed using only arithmetic primitives and indexed reduction, the computation is vectorizable in any numeric library.
Further vectorization over actions is in principle possible, but we defer this to later sections.
The degree of improvement from array vectorization will vary with the language used.
More importantly, it sets the stage for our subsequent innovations.

\subsection{Pre-computed Repetitions}
\label{sec:pre-reps}

The next improvement comes from recognizing that each time we compute a deviation-payoff vector $\devpay(\mix)$, we make use of the $\reps{\prof \mid a}$ value for every profile-action pair.
Since equilibrium search involves many, many deviation-payoff calculations, we can save significant work by pre-computing these repetitions and storing them in a third parallel array.
Then the probability calculation in Equation~\eqref{eq:pi} can extract row $a$ of this repetitions array and element-wise multiply by the array of products.
Our \texttt{gameanalysis.jl} library\footnotemark[1] and the Worked Examples Appendix\footnotemark[5] each include a version of this data structure that computes $\devpay(\mix)$ in a single pass instead of computing $\devpay_a(\mix)$ separately for each action, but we defer the detailed explanation of this vectorization to the following variant where it is dramatically simplified.

\subsection{Opponent Configurations}
\label{sec:opp-config}

A closer inspection of these parallel array structures reveals some redundancies: first the payoffs array contains a number of meaningless entries, since whenever $\prof_a = 0$ the corresponding $\purepay_a$ is masked out of the calculations.\footnotemark[5]
Second, the repetitions array contains a number of redundant entries: whenever two profiles differ by a single player's action,
that is $(\prof \mid a) = (\prof\,' \mid a')$ we will end up storing identical entries for $\reps{\prof \mid a}$ and $\reps{\prof\,' \mid a'}$.

Both of these issues can be avoided if we switch to storing opponent-configurations, with a shared index $j$ over all configurations.
This gives us two parallel arrays of size $A \times \binom{P+A-2}{P-1}$ for configurations and payoffs, and a $1 \times \binom{P+A-2}{P-1}$ repeats-array.
The configurations-array now stores each configuration $\config^j$ over $P-1$ opponents.
Each corresponding column of the repeats-array stores $\reps{\config^j}$, and the same column of the payoffs-array stores, for each action $a$, the payoff $v_a(\config^j)$ for a player choosing $a$ when their opponents play according to $\config^j$.

This also simplifies the deviation-payoffs calculation.
We compute probability $p^j = \Pr(\config^j)$ of a configuration in same way as before, except we can skip the steps from Equations \ref{eq:mi} and \ref{eq:ci}, instead simply accessing configuration $\config^j$ from storage:
\begin{equation*}
    p^j \gets \reps{\config^j} \prod_{a^\prime} \left( \mix_{a^\prime} \right) ^ {\config^{j}_{a^\prime}}
\end{equation*}
The key difference in this equation is that we compute the probability of a configuration $j$, instead of the probability of a configuration derived from a masked profile $i$.
This change removes the need to mask payoffs and means that the configuration probabilities are identical for all deviations, meaning we do not have to perform a separate probability calculation for each action.
This single array of configuration probabilities in turn simplifies vectorization over deviation actions, easily giving us the whole deviation-payoff vector at once.
To get deviation payoffs, we multiply the configuration probabilities $p^j$ by the payoffs-array (broadcasting over actions), and sum over the configurations dimension.
\begin{equation}
    \label{eq:dev_conf}
    \devpay(\mix) \gets \sum_j p^j \vec{\purepay}^j
\end{equation}
Here $\vec{\purepay}^j$ refers to a column of the payoffs-array, and summing these configuration-payoff vectors gives us a vector of deviation payoffs.
The code for this version is the easiest to understand, so despite our recommended further improvements, we include it in the Julia Code Appendix.\footnote{\label{note:julia_code}Implementation shown in Appendix B: Julia Code. \url{https://arxiv.org/abs/2302.13232}}

\subsection{Pre-Weighting by Repetitions}

The next improvement comes from the realization that by associativity we can re-group Equation~\eqref{eq:dev_conf} as follows.

\begin{equation}
\devpay_a(\mix) \gets \sum_{j} \Bigl( \purepay_a \left( \config^j \right) \reps{ \config^j } \Bigr) \prod_{a'} \left( \mix_{a'} \right)^{\config^{j}_{a'}}
\end{equation}

Which means that even though repetitions are logically part of the probability calculation, we can simplify our computations by storing them with the payoff values.
Specifically, we can combine the repeats-array into the payoffs-array by multiplying each payoff value by that configuration's repetitions, so that the entry in row $a$, column $j$ of the payoffs array stores the value:
\begin{equation*}
\reps{ \config^j }  \purepay_a \left( \config^j \right)
\end{equation*}
This operation can be performed once in advance, saving space and speeding up all subsequent deviation payoff calculations.\footnotemark[5]

\subsection{Log Transformation}
\label{sec:log-probs}

The combined effect of all the improvements so far is sufficient to allow reasonably efficient deviation payoff calculations for games with over 100 players, as long as the number of actions is kept small, but this poses a new problem: $\binom{32}{6,6,6,7,7} > 2^{63}$, meaning that the repetitions overflow a 64-bit integer for some profiles with $P=33$ and $A=5$.
We can solve this problem by working with log-probabilities, and this incidentally produces a slight computational speed-up by making use of arithmetic operations that consume fewer processor cycles when we transform exponents into multiplication and multiplication into addition.
Specifically, we can store the natural (or other base) log of the repetition-weighted payoffs $\logrepdevpay$, and to calculate repetitions, we can use a log-gamma function to avoid overflows.

\begin{equation*}
    \logrepdevpay^{j}_{a} \gets \log \left( \reps{\config^j} \right) + \log \left( \purepay_a(\config^j) \right)
\end{equation*}

This will not work if payoffs are negative, but since any positive affine transformation of utilities has no effect on incentives, we can transform any game into one with non-negative payoffs.
In fact, we find it useful under all of our game representations to transform the payoffs into a standardized range, to simplify hyperparameter tuning for various equilibrium-computation algorithms.
And of course these calculations can be vectorized as before with the exponential applied element-wise to an array with contribution-values for each configuration.

\begin{table}[b]
\small
\caption{For a given number of actions $A$, the largest number of players $P$ before \emph{repetitions} overflows a 64-bit integer.}
\begin{tabular}{c|c|c|c|c|c|c|c|c|c|}
\cline{2-10}
$A$ & 2 & 3 & 4 & 5 & 6 & 7 & 8 & 9 & 10 \\ \hline
$\max P$ & 67 & 44 & 36 & 32 & 29 & 27 & 26 & 25 & 25 \\ \cline{2-10}
\multicolumn{9}{c}{} \\ \cline{2-10}
$A$ & 11 & 12 & 13 & 14 & 15 & 16 & 17 & 18 & $\ge \text{19}$ \\ \hline
$\max P$ & 24 & 23 & 23 & 23 & 23 & 22 & 22 & 22 & 21 \\ \cline{2-10}
\end{tabular}
\center
\label{tab:int_overflow}
\end{table}

We can now compute deviation payoffs using this representation by first computing each configuration's contribution $\devcont^j$ to the deviation payoff in log-space.
\begin{equation}
    \label{eq:log_cont}
    \devcont^j \gets \exp \left( \lambda^j + \sum_{a^\prime} \config^{j}_{a^\prime} \log \mix_{a^\prime} \right)
\end{equation}
and then summing over all configurations.\footnotemark[5]
\begin{equation*}
    \devpay_a(\mix) \gets \sum_j \devcont^j
\end{equation*}

\subsection{GPU Acceleration}

Now that our deviation-payoff calculations are done as a sequence of simple mathematical operations broadcast across large arrays of floating-point data, an obvious way to accelerate them is to move the computation to a graphics processor.
Most modern programming languages have libraries that make GPU translation manageable for operations as simple as ours, and specifically in Julia the translation is trivial, requiring only that we choose an array data type that moves the data to the GPU; the code for CPU and GPU versions of our final data structure is otherwise identical.\footnotemark[6]

\subsection{Batch Processing}

The final improvement we implemented requires no change to the preceding data structure, but instead uses it more efficiently.
If memory allows, we can take even greater advantage of SIMD operations on the GPU by computing deviation payoffs for a batch containing multiple mixtures.\footnotemark[6]
This is useful because many of the algorithms we use to compute Nash equilibria are forms of local search where it can help to initialize them from many starting points.
For example, in both replicator dynamics and gain-gradient descent, we generally restart the algorithm as many as 100 times from different initial mixtures, so by parallelizing deviation-payoff calculations, we can parallelize multiple restarts of the algorithm.
This adds an extra dimension to the intermediate arrays constructed by the deviation payoff computation, giving them dimension actions~$\times$ configurations $\times$ mixtures.
This is mainly helpful with small games, since large games can occupy all GPU execution units with the deviation-payoff calculation for a single mixture.
If batch processing is used on large games, the number of mixtures must be carefully chosen to not exceed available GPU memory.

\subsection{Deviation Derivatives}

For any of the data structure variants we describe, it is possible to calculate deviation derivatives, but here we describe the approach only for the final log-transformed variants.

Computing the Jacobian with respect to the mixture probability $\mix_s$ only requires a small addition to the deviation payoff contributions computed in equation~\eqref{eq:log_cont}; we multiply by $\frac{\config_s}{\mix_{s}}$ 
over a new dimension $s$, before summing across configurations.\footnotemark[6]
\begin{equation*}
    \devderiv{\mix}{a}{\mix_s} \gets \sum_j \frac{\config^{j}_{s}}{\mix_{s}} \devcont^j
\end{equation*}

This calculation can be fully vectorized using similar operations but adds a dimension since we have a partial derivative for each \emph{pair} of actions.
This extra dimension is the main reason gradient descent runs slower than replicator dynamics in our experiments.

\section{Validation and Experiments}


\subsection{Size Limits}
\label{sec:sizes}

\begin{figure}[t]
\center
\includegraphics[width=\columnwidth]{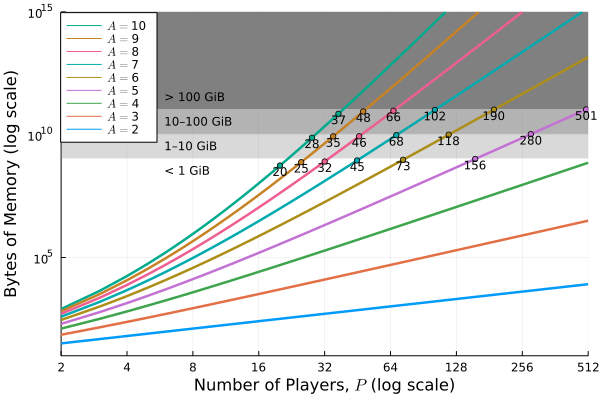}
\caption{GPU memory required to store the final version of our data structure. Batch-computation of deviation payoffs increases memory footprint by a linear factor.}
\label{fig:mem_limit}
\end{figure}

The first constraint on the size of the data structures comes from integer overflows when calculating repetitions.
Table~\ref{tab:int_overflow} shows the upper limit in terms of $P$ and $A$ beyond which at least one profile will overflow a signed 64-bit integer.
For example, if we want to represent a 5-action game with more than 32 players, we must use the log-transform to ensure that we calculate probabilities correctly.
Note that using unsigned integers usually makes no difference and never allows more than one extra player.

This raises the question of whether we sacrifice precision when using the log-transform.
To test this, we generated 10 random bipartite action-graph games for every combination of $2 \le P \le 512$ and $2 \le A \le 20$ for which the size of our data structures would be below 1GiB.
We then represented these games using the CPU (64-bit) and GPU (32-bit) versions of our data structure, and calculated deviation payoffs for 1000 mixtures per game and recorded the largest error in the calculated deviation payoffs.
These errors were tiny, except in games with hundreds of players in the 32-bit representation, and even then the scale of the errors was at least 7 orders of magnitude smaller than the payoff range of the game.
Full results are shown in the Supplementary Figures Appendix.\footnote{\label{note:supplementary_figures}See Appendix D: supplementary figures. \url{https://arxiv.org/abs/2302.13232}}

With these results, the question of what games we can represent comes down to what can fit in (main or GPU) memory.
Figure~\ref{fig:mem_limit} shows the size of the GPU data structure as $P$ increases for various values of $A$.
A similar figure for 64-bit arrays appears in Appendix~D.\footnotemark[7]
Note that calculating deviation payoffs requires storing intermediate arrays of similar size to the data structure, so game size should be restricted to below half of available memory (or smaller if we want to calculate deviation payoffs in batches).
Note that these sizes mean that it is entirely reasonable to store a 100-player, 6-action game.

\subsection{Timing Comparisons}
\label{sec:timing}

\begin{figure}
\center
\includegraphics[width=\columnwidth]{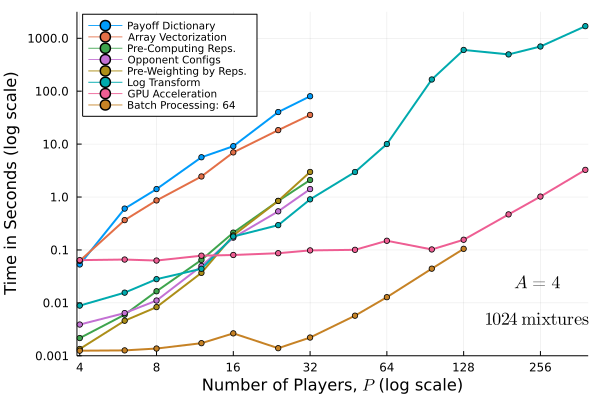}
\caption{Time to required to compute deviation payoffs for 1024 mixtures in 4-action games using each data structure variant. Lines stop when either an integer overflow is encountered or when more than 1GB of memory is required. Similar plots for $A=6$ and $A=8$ appear in Appendix D.}
\label{fig:dev_pay_timing}
\end{figure}

\begin{figure}[t]
\center
\includegraphics[width=\columnwidth]{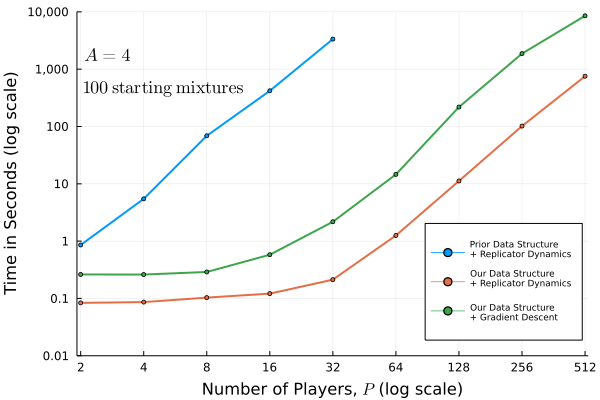}
\caption{Time to compute Nash equilibria in 4-action games with 100 starting mixtures and 1000 iterations. Replicator dynamics is considerably faster than gradient descent (and could get away with fewer iterations), but it is often worthwhile to run both. Our data structure lets us represent and solve much larger games than the previous state-of-the-art. Similar plots for $A=6$ and $A=8$ appear in Appendix D.}
\label{fig:eq_timing}
\end{figure}

Figures \ref{fig:dev_pay_timing} and \ref{fig:eq_timing} show our main results: our data structures produce significant speedup in computing deviation payoffs and therefore running Nash-finding algorithms.
Timing comparisons were run using an AMD Threadripper 2990WX CPU and NVIDIA RTX 2080-Ti GPUs.
Figure~ \ref{fig:dev_pay_timing} shows the time to compute deviation payoffs using each of the data structures described in the previous section.
The first several lines end at $P=32$, since this is near the int-overflow limit for $A=4$.
The batch-processing line ends at $P=128$, because operating on a batch of 64 mixtures requires much more memory.
Note that our proposed data structure can compute deviation payoffs in a game with 384 players faster than the baseline data structure can handle $P=12$.
Many more deviation-payoff timing comparisons including 6- and 8-action games appear in Appendix~D.\footnotemark[7]

Figure~\ref{fig:eq_timing} shows a similar result for Nash-finding, but focusing only on our best data structure, and using batch sizes adapted to available memory.
First, note that replicator dynamics using our data structure outperforms the previous implementation by four orders of magnitude.
Gradient descent is consistently an order of magnitude slower than replicator dynamics, making it impractical and not generally implemented with older data structures.
But with our data structure that slowdown can often be acceptable because gradient descent and replicator dynamics frequently identify some distinct equilibria (as illustrated in Figure~\ref{fig:nash_traces}), and so in cases where replicator dynamics fails or more equilibria are desired it is entirely reasonable to run both algorithms.
For example, with $A=4$ and $P=512$, we can perform 100 runs of replicator dynamics or gradient descent in a reasonable time-frame.
Timing results shown in Appendix~D for $A=6$ and $A=8$ are broadly similar.\footnotemark[7]

\section{Extensions}

\subsection{Many Strategies}
\label{sec:many-strategies}

The size of our data structures and therefore the time to compute deviation payoffs scales reasonably similarly with respect to the number of players $P$ and the number of actions $A$.
The configurations and payoffs arrays both have size $A \times \binom{P+A-2}{P-1}$, which is dominated by the binomial that grows exponentially in the smaller of $P$ and $A$.
As a result, it is generally possible to represent games with a large number of players but a small number of actions or a large number of actions but a small number of players, with the caveat that actions contribute an extra linear factor.

Unfortunately, equilibrium computation does not scale equally well in players and actions.
Identifying symmetric mixed-strategy Nash equilibria means searching for specific points in the mixed-strategy simplex, whose dimension grows with the number of actions.
The various complete algorithms are generally hopelessly slow with large numbers of actions, and the local search methods we prefer are much less likely to identify an equilibrium on any given run.
This means that solving symmetric games with many players is far more feasible than solving ones with many actions.

To handle games with many actions, we can take inspiration from \citet{simple_search}, and search for equilibria with small support sets.
This requires solving a subgame defined by restricting the game to a subset of the actions, and then checking whether those subgame solutions have beneficial deviations in the full action set.
To facilitate this, we can use a variant of our data structure where we include additional rows in the payoffs-array representing actions outside the support set, but make no change to the configurations-array.
Then for a mixture in the support set, we can compute deviation payoffs for all actions against that mixture by expanding the step where we broadcast the multiplication of configuration probabilities and payoffs.
If we determine that the current subgame set does not contain a Nash equilibrium, we can update the support set replacing weaker strategies with stronger ones, as measured by deviation gain against the subgame's candidate solutions (or other metrics).
This will iteratively search the space of supports and eventually identify a small-support equilibrium if any exist.
This approach of constructing complete subgames with extra payoff information for unilateral deviations is already widely used in empirical game settings \cite{EGTA} where determining payoffs for a single configuration is expensive.

\subsection{Role-Symmetric Games}

\label{sec:role-symmetry}

Outside of fully-symmetric games, many large normal-form games exhibit role-symmetry, where players can be grouped into $R$ roles---like buyers and sellers or attackers and defenders---and within each role, all players are indistinguishable.
To solve role symmetric games, analysts typically search for role-symmetric Nash equilibria, where all players in a given role play the same mixed strategy.
Role symmetric mixed-strategy profiles can be represented by a vector that concatenates the mixed strategy played by each role, and the deviation payoff vector of the same dimension gives the expected utility of a unilateral deviator selecting each action.
These deviation payoffs and their derivatives remain central to computing role-symmetric equilibria, and almost all of the techniques we propose for improving representations of symmetric games also apply under role symmetry.

Unfortunately, representing opponent configurations gets trickier with more than one role, because the set of opponents is different for players belonging to different roles.
For example, in a game with 10 buyers and 10 sellers, when computing deviation payoffs for buyer-actions, opponent configurations include 9 buyers and 10 sellers, whereas for seller actions, there are 10 buyer-opponents and 9 seller-opponents.
We could resolve this by building separate opponent-configuration tables for each role $r$, where the arrays for role $r$ are based on configurations with $P_r - 1$ opponents in the same role and $P_{r'}$ opponents in other roles.
This gives us $R$ pairs of configuration- and payoff-arrays where the arrays for role $r$ have $\binom{P_{r} + A_{r} - 2}{P_{r}-1} \left(\prod_{r' \ne r} \binom{P_{r'} + A_{r'} - 1}{P_{r'}} \right) $ columns, by $\left(\sum_{r=1}^{R} A_r \right)$ rows for configurations and $A_r$ rows for payoffs.

We could also dispense with the opponent-configuration approach and instead store full $P$-player profiles.
This results in one profile-array and one payoff-array that each have size $ \left(\sum_{r=1}^{R} A_r \right) \times \left(\prod_{r=1}^{R} \binom{P_r + A_r - 1}{P_r} \right) $, but requires us to return to masking-based dev\-iation-payoff computations.
Thus for multi-role games, we have a choice between slightly smaller storage requirements for a profile-based data structure or slightly simpler computations for a config\-uration-based representation;
the \texttt{gameanalysis.py} library\footnotemark[2] employs the former option.
Under either of these approaches, all of our other optimizations can still apply, and as long as the number of roles is small both options provide for reasonably efficient deviation-payoff calculations.

\subsection{Action-Graph Games}

Action-graph games \cite{action-graph_games} represent certain games compactly by storing, for each node, a mapping from neighborhood configurations to payoffs.
Each action (for any player) corresponds to a node in a graph, and an action's payoff depends only on the number of players choosing each of the adjacent actions.
This means that action-graph games are role-symmetric, with any group of players who share an action set belonging to the same role; when all players have the same action set, action-graph games are symmetric.

In the symmetric case, we can extend our data structures to efficiently compute deviation payoffs in action-graph games with a pair of arrays for each action storing that action's neighborhood configurations and repetition-weighted payoffs.
The key augmentation is to include, as part of the representation of an opponent-configuration, an extra ``action'' capturing the number of players choosing any action outside the neighborhood.
Then when computing the deviation payoff for an action, we can sum the probabilities for each non-adjacent action to get the out-of-neighborhood probability.
This means that all of our data structure improvements can be applied to symmetric AGG-$\emptyset$ games.

For role-symmetric AGG-$\emptyset$ games, our role-symmetric data structure variants could be applied after splitting up action nodes shared by more than one role.
For action-graph games with contribution-independent function nodes, many of our data-structure improvements can be applied, but the representation of these games tends to be sufficiently compact that the potential gains may be small.
The \texttt{gameanalysis.jl} library\footnotemark[1] implements symmetric action-graph games with a bipartite graph between actions and contribution-independent functions as one of the tools for generating interesting random symmetric game instances.
This implementation makes use of vectorization, opponent configurations, pre-computed repetitions, and the log transformation.

\subsection{Multi-Threading}

The next extension we would like to explore would focus on improving the CPU version of our data structure to capture some of the parallelism achievable on the GPU.
At present, our deviation-payoff calculations are single-threaded, and while we might not expect to outperform SIMD array operations, there should still be room for significant speedup on multi-core systems from parallelizing a batch of deviation payoff computations across threads, especially for games in the multi-gigabyte size-range that can stress GPU memory but still fit in system RAM.

\section{Conclusion}

Our validation experiments show that the data structures we propose are capable of storing (Section~\ref{sec:sizes}) and efficiently computing (Section~\ref{sec:timing}) deviation payoffs in symmetric games with very large numbers of players.
By using incomplete-but-effective search algorithms we are consistently able to identify symmetric Nash equilibria.
In combination with the iterative exploration approach described in Section~\ref{sec:many-strategies} we can also find small-support equilibria in games with many actions.
These results dramatically out-class existing solvers, enabling analysis of a much wider range of symmetric games, and closing the gap between micro- and macro-scale symmetric games.

\balance
\bibliographystyle{ACM-Reference-Format} 
\bibliography{dev_pays.bib}

\vfill

\pagebreak
\appendix
\onecolumn


\section{Appendix A: Worked Examples}

We will illustrate the various iterations of our data structure and the corresponding deviation payoff calculations using the following example of a 3-player, 3-action symmetric game.
In this game, each player selects an action $a \in \{1,2,3\}$, and if a player's action $a$ is unique, their payoff is $a$.
If two players select action $a$, they both receive payoff $-a$; if all three players select $a$, they each receive 0.
The standard normal-form representation of this game is shown in Figure~\ref{fig:example-NFG}.

\begin{figure}[ht]
\center
\begin{tabular}{c|c|c|c|c}
\multicolumn{1}{c}{} & \multicolumn{1}{c}{\bf 1} & \multicolumn{1}{c}{\bf 2} & \multicolumn{1}{c}{\bf 3} \\ \cline{2-4}
\bf 1 & $0,0,0$ & $-1,2,-1$ & $-1,3,-1$ \\ \cline{2-4}
\bf 2 & $2,-1,-1$ & $-2,-2,1$ & $2,3,1$ & \bf 1 \\ \cline{2-4}
\bf 3 & $3,-1,-1$ & $3,2,1$ & $-3,-3,1$ \\ \cline{2-4}
\end{tabular}

\vspace{2mm}
\begin{tabular}{c|c|c|c|c}
\multicolumn{1}{c}{} & \multicolumn{1}{c}{\bf 1} & \multicolumn{1}{c}{\bf 2} & \multicolumn{1}{c}{\bf 3} \\ \cline{2-4}
\bf 1 & $-1,-1,2$ & $1,-2,-2$ & $1,3,2$ \\ \cline{2-4}
\bf 2 & $-2,1,-2$ & $0,0,0$ & $-2,3,-2$ & \bf 2 \\ \cline{2-4}
\bf 3 & $3,1,2$ & $3,-2,-2$ & $-3,-3,2$ \\ \cline{2-4}
\end{tabular}

\vspace{2mm}
\begin{tabular}{c|c|c|c|c}
\multicolumn{1}{c}{} & \multicolumn{1}{c}{\bf 1} & \multicolumn{1}{c}{\bf 2} & \multicolumn{1}{c}{\bf 3} \\ \cline{2-4}
\bf 1 & $-1,-1,3$ & $1,2,3$ & $1,-3,-3$ \\ \cline{2-4}
\bf 2 & $2,1,3$ & $-2,-2,3$ & $2,-3,-3$ & \bf 3 \\ \cline{2-4}
\bf 3 & $-3,1,-3$ & $-3,2,-3$ & $0,0,0$ \\ \cline{2-4}
\end{tabular}

\caption{Normal-form representation of our example 3-player, 3-action symmetric game.}
\label{fig:example-NFG}

\end{figure}

\noindent
For each data structure, we will show the calculation of $\devpay_1(\mix)$ for $\mix = \angles{0.1, 0.5, 0.4}$, that is the deviation payoff of action 1 when both opponents play a mixture that places a 10\% probability on 1, a 50\% probability on 2 and a 40\% probability on 3.
In cases where the computation naturally vectorizes over actions, we will show how to compute $\devpay(\mix)$.

\subsection{Payoff Dictionary}

If we represent our example game using a mapping from profiles to payoffs, the mapping will store the following $\binom{P+A-1}{P} = 10$ pairs.
We denote extraneous entries in the payoff vector by a dot.

\begin{align*}
\threevec{3}{0}{0} &\;\rightarrow\; \threevec{0}{\cdot}{\cdot}		& \threevec{2}{1}{0} &\;\rightarrow\; \threevec{-1}{2}{\cdot}
	& \threevec{2}{0}{1} &\;\rightarrow\; \threevec{-1}{\cdot}{3}	& \threevec{1}{2}{0} &\;\rightarrow\; \threevec{1}{-2}{\cdot} 
	& \threevec{1}{1}{1} &\;\rightarrow\; \threevec{1}{2}{3} \\
\threevec{1}{0}{2} &\;\rightarrow\; \threevec{1}{\cdot}{-3}		& \threevec{0}{3}{0} &\;\rightarrow\; \threevec{\cdot}{0}{\cdot}
	& \threevec{0}{2}{1} &\;\rightarrow\; \threevec{\cdot}{-2}{3}	& \threevec{0}{1}{2} &\;\rightarrow\; \threevec{\cdot}{2}{-3}
	& \threevec{0}{0}{3} &\;\rightarrow\; \threevec{\cdot}{\cdot}{0}
\end{align*}

\noindent
To calculate $\devpay_1(\mix)$, we begin from Equation~\eqref{eq:prof_dev_pays}, giving

\begin{align*}
\devpay_1(\mix) \;=\; &\sum_{\prof \mid \prof_1 > 0} \Pr_{\mix} \left( \prof \mid 1 \right) \purepay_1(\prof) \\ \\
=\; &\Pr( \angles{3,0,0}) \cdot \purepay_1(\angles{3,0,0}) \; +
	\Pr( \angles{2,1,0}) \cdot \purepay_1(\angles{2,1,0}) \; +
	\Pr( \angles{2,0,1}) \cdot \purepay_1(\angles{2,0,1}) \; + \\
&\;\;\;\;\Pr( \angles{1,2,0}) \cdot \purepay_1(\angles{1,2,0}) \; +
	\Pr( \angles{1,1,1}) \cdot \purepay_1(\angles{1,1,1}) \; +
	\Pr( \angles{1,0,2}) \cdot \purepay_1(\angles{1,0,2}) \\ \\
=\; & (1) (0.1^2) (0) \;+
	(2) (0.1 \cdot 0.5) (-1) \;+
	(2) (0.1 \cdot 0.4) (-1) \;+ \\
&\;\;\;\;(1) (0.5^2) (1) \;+
	(2) (0.5 \cdot 0.4) (1) \;+
	(1) (0.4^2) (1) \\ \\
=\; &0.63
\end{align*}

\noindent
So this computation loops over all six of the profiles where action 1 is played.
Calculating $\devpay_2(\mix)$ and $\devpay_3(\mix)$ requires similar loops over different subsets of the profiles.

\subsection{Array Vectorization}

If we use parallel arrays to store the profiles and payoffs, our example game is stored in the following $3 \times 10$ arrays.
Note that each column of the tables corresponds to a profile/payoff pair from the dictionary version.\\

\begin{minipage}{.5\textwidth}
\noindent
Profiles:\\
\texttt{
\begin{tabular}{|C{3mm}|C{3mm}|C{3mm}|C{3mm}|C{3mm}|C{3mm}|C{3mm}|C{3mm}|C{3mm}|C{3mm}|}
\hline
3 & 2 & 2 & 1 & 1 & 1 & 0 & 0 & 0 & 0 \\ \hline
0 & 1 & 0 & 2 & 1 & 0 & 3 & 2 & 1 & 0 \\ \hline
0 & 0 & 1 & 0 & 1 & 2 & 0 & 1 & 2 & 3 \\ \hline
\end{tabular}
}
\end{minipage}
\begin{minipage}{.5\textwidth}
\noindent
Payoffs:\\
\texttt{
\begin{tabular}{|C{3mm}|C{3mm}|C{3mm}|C{3mm}|C{3mm}|C{3mm}|C{3mm}|C{3mm}|C{3mm}|C{3mm}|}
\hline
\tt
0 & -1 & -1 & 1 & 1 & 1 & $\cdot$ & $\cdot$ & $\cdot$ & $\cdot$ \\ \hline
$\cdot$ & 2 & $\cdot$ & -2 & 2 & $\cdot$ & 0 & -2 & 2 & $\cdot$ \\ \hline
$\cdot$ & $\cdot$ & 3 & $\cdot$ & 3 & -3 & $\cdot$ & 3 & -3 & 0 \\ \hline
\end{tabular}
}
\end{minipage}\\

\noindent
The deviation-payoff calculation $\devpay_1(\mix)$ is fundamentally the same as in the previous section, but performed with element-wise array operations instead of explicit loops.
For clarity, we will show some of the intermediate steps here.
First, we mask out the entries where action 1 is not played, decrement the count for action 1, and use the resulting profiles to exponentiate the probabilities from $\mix$.\\

\noindent
\begin{tabular}{|C{2mm}|}
\hline
.1 \\ \hline
.5 \\ \hline
.4 \\ \hline
\end{tabular}
$\bigdot^{\wedge}$
\begin{tabular}{|C{1mm}|C{1mm}|C{1mm}|C{1mm}|C{1mm}|C{1mm}|}
\hline
2 & 1 & 1 & 0 & 0 & 0 \\ \hline
0 & 1 & 0 & 2 & 1 & 0 \\ \hline
0 & 0 & 1 & 0 & 1 & 2 \\ \hline
\end{tabular}
$=$
\begin{tabular}{|C{1.5mm}|C{1.5mm}|C{1.5mm}|C{1.5mm}|C{1.5mm}|C{2mm}|}
\hline
\!\!.01 & .1 & .1 & 1 & 1 & 1 \\ \hline
1 & .5 & 1 & \!\!.25 & .5 & 1 \\ \hline
1 & 1 & .4 & 1 & .4 & \!\!.16 \\ \hline
\end{tabular}\\

\noindent
Where the $\bigdot^{\wedge}$ operator indicates element-wise exponentiation broadcast across the rows: 
that is, element $r,c$ of the output matrix comes from raising element $r$ of the input vector to the power of element $r,c$ in the input matrix.
Then we can get the probability of each profile by taking a product down each column and multiplying element-wise by an array of repeats, which we compute by applying a multinomial function to each column of the same masked-and-decremented profiles array.\\

\begin{tabular}{|C{1.5mm}|C{1.5mm}|C{1.5mm}|C{1.5mm}|C{1.5mm}|C{1.5mm}|}
\hline
\!\!.01 & \!\!.05 & \!\!.04 & \!\!.25 & .2 & \!\!.16 \\ \hline
\multicolumn{6}{c}{$\bigdot\times$} \\ \hline
1 & 2 & 2 & 1 & 2 & 1 \\ \hline
\end{tabular}
$\;=\;$
\begin{tabular}{|C{1.5mm}|C{1.5mm}|C{1.5mm}|C{1.5mm}|C{1.5mm}|C{1.5mm}|}
\hline
\!\!.01 & .1 & \!\!.08 & \!\!.25 & .4 & \!\!.16 \\ \hline
\end{tabular}\\

\noindent
We can then apply the same mask to the payoffs array, slice out the row for action 1, element-wise multiply with the profile-probabilities array, and sum to get the deviation payoff.

\begin{minipage}{.045\columnwidth}
\vfill
\[ \sum \Biggl( \]
\vfill
\end{minipage}
\begin{tabular}{|C{3mm}|C{3mm}|C{3mm}|C{3mm}|C{3mm}|C{3mm}|}
\hline
\!\!.01 & .1 & \!\!.08 & \!\!.25 & .4 & \!\!.16 \\ \hline
\multicolumn{6}{c}{$\bigdot\times$} \\ \hline
0 & -1 & -1 & 1 & 1 & 1 \\ \hline
\end{tabular}
$\Biggr)\;=\; 0.63$

\subsection{Pre-computed Repetitions}

Our next improvement came from pre-computing $\reps{\prof \mid a}$ for each profile-action pair, and storing them in a third parallel array.
Here is the resulting array for our example game.\\

\noindent
Repeats:\\
\texttt{
\begin{tabular}{|C{3mm}|C{3mm}|C{3mm}|C{3mm}|C{3mm}|C{3mm}|C{3mm}|C{3mm}|C{3mm}|C{3mm}|}
\hline
1 & 2 & 2 & 1 & 2 & 1 & $\cdot$ & $\cdot$ & $\cdot$ & $\cdot$ \\ \hline
$\cdot$ & 1 & $\cdot$ & 2 & 2 & $\cdot$ & 1 & 2 & 1 & $\cdot$ \\ \hline
$\cdot$ & $\cdot$ & 1 & $\cdot$ & 2 & 2 & $\cdot$ & 1 & 2 & 1 \\ \hline
\end{tabular}
}\\

\noindent
The deviation-payoff calculation for a single action using this version would identical to the one just described,
except that this array of configuration-repetitions:\\

\noindent
\begin{tabular}{|C{1.5mm}|C{1.5mm}|C{1.5mm}|C{1.5mm}|C{1.5mm}|C{1.5mm}|}
\hline 1 & 2 & 2 & 1 & 2 & 1 \\ \hline
\end{tabular}\\

\noindent
would now be simply a masked slice out of the repeats-array, avoiding the need to re-calculate multinomials for every deviation-payoff computation.
But we also discussed how to compute the full vector of deviation payoffs $\devpay(\mix)$ at once.
The computation begins by raising the mixture probabilities to each of the profile counts.\\

\noindent
\begin{tabular}{|C{2mm}|}
\hline
.1 \\ \hline
.5 \\ \hline
.4 \\ \hline
\end{tabular}
$\bigdot^{\wedge}$
\begin{tabular}{|C{1mm}|C{1mm}|C{1mm}|C{1mm}|C{1mm}|C{1mm}|C{1mm}|C{1mm}|C{1mm}|C{1mm}|}
\hline
3 & 2 & 2 & 1 & 1 & 1 & 0 & 0 & 0 & 0 \\ \hline
0 & 1 & 0 & 2 & 1 & 0 & 3 & 2 & 1 & 0 \\ \hline
0 & 0 & 1 & 0 & 1 & 2 & 0 & 1 & 2 & 3 \\ \hline
\end{tabular}
$\;\;=\;\;$
\begin{tabular}{|C{2.5mm}|C{2mm}|C{2mm}|C{2mm}|C{2mm}|C{2mm}|C{2.5mm}|C{2mm}|C{2mm}|C{2.5mm}|}
\hline
\!\!\!.001 & \!\!.01 & \!\!.01 & .1 & .1 & .1 & 1 & 1 & 1 & 1 \\ \hline
1 & .5 & 1 &  \!\!.25 & .5 & 1 & \!\!\!.125 &  \!\!.25 & .5 & 1 \\ \hline
1 & 1 & .4 & 1 & .4 & \!\!.16 & 1 & .4 & \!\!.16 & \!\!\!.064 \\ \hline
\end{tabular}\\

\noindent
We can then take a product down each column, giving an array of the $\prod_a$ terms from Equation~\eqref{eq:prof_prob}.
We then want to multiply this by the repeats-array (with zeros filled in for the missing elements) element-wise divided by the mixture probabilities:\\

\noindent
\begin{tabular}{|C{3.5mm}|C{3.5mm}|C{3.5mm}|C{3.5mm}|C{3.5mm}|C{3.5mm}|C{3.5mm}|C{3.5mm}|C{3.5mm}|C{3.5mm}|}
\hline
\!\!.001 & \!\!.005 & \!\!.004 & \!\!.025 & \!.02 & \!\!.016 & \!\!.125 & .1 & \!.08 & \!\!.064 \\ \hline
\multicolumn{10}{c}{$\bigdot\times$} \\ \hline
10 & 20 & 20 & 10 & 20 & 10 & 0 & 0 & 0 & 0 \\ \hline
0 & 2 & 0 & 4 & 4 & 0 & 2 & 4 & 2 & 0 \\ \hline
0 & 0 & \!2.5 & 0 & 5 & 5 & 0 & \!2.5 & 5 & \!2.5 \\ \hline
\multicolumn{10}{c}{=} \\ \hline
\!.01 & .1 & \!.08 & \!.25 & .4 & \!.16 & 0 & 0 & 0 & 0 \\ \hline
0 & \!.01 & 0 & .1 & \!.08 & 0 & \!.25 & \!.4 & \!.16 & 0 \\ \hline
0 & 0 & \!.01 & 0 & .1 & .08 & 0 & \!.25 & .4 & \!.16 \\ \hline
\end{tabular}\\

\noindent
Note that the first six entries in the top row of this array correspond to the $\Pr_{\mix}(\prof \mid 1)$ array from before.
Likewise, the non-zero entries in the second and third rows give the various $\Pr_{\mix}(\prof \mid 2)$ and $\Pr_{\mix}(\prof \mid 3)$ probabilities.
We can now element-wise multiply by the payoffs array (again filling in zeros for missing entries), then sum across the rows to get the vector of deviation payoffs.\\

\noindent
\begin{minipage}{.05\textwidth}
\vfill
\[ \sum_{rows} \Biggl( \]
\vfill
\end{minipage}
\begin{minipage}{.315\textwidth}
\begin{tabular}{|C{1mm}|C{3mm}|C{4mm}|C{3mm}|C{3mm}|C{4mm}|C{1mm}|C{3mm}|C{4mm}|C{1mm}|}
\hline
0 & \!\!-.1 & \!\!\!-.08 & \!.25 & .4 & \!.16 & 0 & 0 & 0 & 0 \\ \hline
0 & \!.02 & 0 & \!\!-.2 & \!.16 & 0 & 0 & \!\!-.8 & \!.32 & 0 \\ \hline
0 & 0 & \!.03 & 0 & .3 & \!\!\!-.24 & 0 & \!.75 & \!\!\!-1.2 & 0 \\ \hline
\end{tabular}
\end{minipage}
\begin{minipage}{.3\textwidth}
\[\Biggr) \;\;=\;\; \threevec{0.63}{-0.5}{-0.36} \]
\end{minipage}

\subsection{Opponent Configurations}

If we move to storing arrays based on configurations over $P-1$ opponents instead of profiles over $P$ players, we shrink our arrays to $\binom{P+A-2}{P-1} = 6$ columns, and can eliminate the actions dimension from the repeats-array.
The resulting tables for our example game are as follows.

\begin{minipage}{.33\textwidth}
\vspace{2mm}
\noindent
Configurations:\\
\texttt{
\begin{tabular}{|C{3mm}|C{3mm}|C{3mm}|C{3mm}|C{3mm}|C{3mm}|}
\hline
2 & 1 & 1 & 0 & 0 & 0 \\ \hline
0 & 1 & 0 & 2 & 1 & 0 \\ \hline
0 & 0 & 1 & 0 & 1 & 2 \\ \hline
\end{tabular}
}
\end{minipage}
\begin{minipage}{.33\textwidth}
\noindent
Payoffs:\\
\texttt{
\begin{tabular}{|C{3mm}|C{3mm}|C{3mm}|C{3mm}|C{3mm}|C{3mm}|}
\hline
0 & -1 & -1 & 1 & 1 & 1 \\ \hline
2 & -2 & 2 & 0 & -2 & 2 \\ \hline
3 & 3 & -3 & 3 & -3 & 0 \\ \hline
\end{tabular}
}
\end{minipage}
\begin{minipage}{.33\textwidth}
\noindent
Repeats:\\
\texttt{
\begin{tabular}{|C{3mm}|C{3mm}|C{3mm}|C{3mm}|C{3mm}|C{3mm}|}
\hline
1 & 2 & 2 & 1 & 2 & 1 \\ \hline
\end{tabular}
}
\end{minipage}\\

\noindent
With this representation, computing a vector of deviation payoffs is simplified considerably.
We begin once again by raising mixture probabilities to the configuration counts.\\

\noindent
\begin{tabular}{|C{2mm}|}
\hline
.1 \\ \hline
.5 \\ \hline
.4 \\ \hline
\end{tabular}
$\bigdot^{\wedge}$
\begin{tabular}{|C{1mm}|C{1mm}|C{1mm}|C{1mm}|C{1mm}|C{1mm}|}
\hline
2 & 1 & 1 & 0 & 0 & 0 \\ \hline
0 & 1 & 0 & 2 & 1 & 0 \\ \hline
0 & 0 & 1 & 0 & 1 & 2 \\ \hline
\end{tabular}
$=$
\begin{tabular}{|C{1.5mm}|C{1.5mm}|C{1.5mm}|C{1.5mm}|C{1.5mm}|C{2mm}|}
\hline
\!\!.01 & .1 & .1 & 1 & 1 & 1 \\ \hline
1 & .5 & 1 & \!\!.25 & .5 & 1 \\ \hline
1 & 1 & .4 & 1 & .4 & \!\!.16 \\ \hline
\end{tabular}\\

\noindent
Then we take the product down each column and multiply column-wise by the repeats-array.\\

\begin{tabular}{|C{1.5mm}|C{1.5mm}|C{1.5mm}|C{1.5mm}|C{1.5mm}|C{1.5mm}|}
\hline
\!\!.01 & \!\!.05 & \!\!.04 & \!\!.25 & .2 & \!\!.16 \\ \hline
\multicolumn{6}{c}{$\bigdot\times$} \\ \hline
1 & 2 & 2 & 1 & 2 & 1 \\ \hline
\end{tabular}
$\;=\;$
\begin{tabular}{|C{1.5mm}|C{1.5mm}|C{1.5mm}|C{1.5mm}|C{1.5mm}|C{1.5mm}|}
\hline
\!\!.01 & .1 & \!\!.08 & \!\!.25 & .4 & \!\!.16 \\ \hline
\end{tabular}\\

\noindent
Note that while the steps so far appear identical to those given for payoff-arrays above, they are simpler because no masking is required in the configurations array and the repeats-array has been pre-computed.
From here, we column-wise multiply by the payoffs array and then sum across rows.\\

\begin{minipage}{.35\textwidth}
\begin{tabular}{|C{4.5mm}|C{4.5mm}|C{4.5mm}|C{4.5mm}|C{4.5mm}|C{4.5mm}|}
\hline
.01 & .1 & .08 & .25 & .4 &.16 \\ \hline
\multicolumn{6}{c}{$\bigdot\times$} \\ \hline
0 & -1 & -1 & 1 & 1 & 1 \\ \hline
2 & -2 & 2 & 0 & -2 & 2 \\ \hline
3 & 3 & -3 & 3 & -3 & 0 \\ \hline
\multicolumn{6}{c}{$=$} \\ \hline
0 & -.1 & \!\!-.08 & .25 & .4 & .16 \\ \hline
.02 & -.2 & \!.16 & 0 & -.8 & .32 \\ \hline
.03 & .3 & \!\!-.24 & .75 & \!\!-1.2 & 0 \\ \hline
\end{tabular}
\end{minipage}
\begin{minipage}{.25\textwidth}
\vspace{24mm}
\[ \sum_{rows} \left( \; \boldsymbol{\cdot} \; \right) = \threevec{0.63}{-0.5}{-0.36} \]
\end{minipage}

\subsection{Weighted Payoffs}

The next trick was to avoid storing the repeats-array by re-arranging the multiplication.
By associativity, we can multiply the repeats by the payoffs instead of by the probabilities.
This can be done in advance, with the result that we store the following weighted payoffs.\\

\noindent
Weighted Payoffs:\\
\texttt{
\begin{tabular}{|C{3mm}|C{3mm}|C{3mm}|C{3mm}|C{3mm}|C{3mm}|}
\hline
0 & -2 & -2 & 1 & 2 & 1 \\ \hline
2 & -4 & 4 & 0 & -4 & 2 \\ \hline
3 & 6 & -6 & 3 & -6 & 0 \\ \hline
\end{tabular}
}\\

\noindent
The deviation payoff calculation proceeds identically to above, except that we can skip one step and the final operation before the sum becomes:\\

\begin{tabular}{|C{4.5mm}|C{4.5mm}|C{4.5mm}|C{4.5mm}|C{4.5mm}|C{4.5mm}|}
\hline
.01 & .1 & .08 & .25 & .4 & .16 \\ \hline
\multicolumn{6}{c}{$\bigdot\times$} \\ \hline
0 & -2 & -2 & 1 & 2 & 1 \\ \hline
2 & -4 & 4 & 0 & -4 & 2 \\ \hline
3 & 6 & -6 & 3 & -6 & 0 \\ \hline
\multicolumn{6}{c}{$=$} \\ \hline
0 & -.1 & \!\!-.08 & .25 & .4 & .16 \\ \hline
.02 & -.2 & \!.16 & 0 & -.8 & .32 \\ \hline
.03 & .3 & \!\!-.24 & .75 & \!\!-1.2 & 0 \\ \hline
\end{tabular}

\subsection{Log Probabilities}

The last major change to the data structure allows us to work with log-probabilities.
Unfortunately, we cannot simply log-transform the weighted payoffs array, because it often has negative entries.
However, since a positive affine transformation of a game's payoffs has no effect on incentives, we can re-scale the payoffs to permit this transformation.
Somewhat arbitrarily, we chose to re-scale the payoffs so that the minimum value becomes $10^{-5}$ and the maximum value becomes $10^3$.
Applying this transformation to the opponent-configuration version of the payoff array gives:\\

\noindent
Scaled Payoffs:\\
\texttt{ \small
\begin{tabular}{|C{9.8mm}|C{9.8mm}|C{9.8mm}|C{9.8mm}|C{9.8mm}|C{9.8mm}|}
\hline
500.00 & 333.33 & 333.33 & 666.67 & 666.67 & 666.67 \\ \hline
833.33 & 166.67 & 833.33 & 500.00 & 166.67 & 833.33 \\ \hline
1000.0 & 1000.0 & 1.0e-5 & 1000.0 & 1.0e-5 & 500.00 \\ \hline
\end{tabular}
}\\

\noindent
We can then multiply by the repeats and take the log-transform, resulting in the following array, which is stored alongside the opponent-configurations array.

\vspace{2mm}
\noindent
Weighted Log-Payoffs:\\
\texttt{ \small
\begin{tabular}{|C{9mm}|C{9mm}|C{9mm}|C{9mm}|C{9mm}|C{9mm}|}
\hline
 6.215 & 6.502 & 6.502 & 6.502 & 7.195 & 6.502 \\ \hline
 6.725 & 5.809 & 7.419 & 6.215 & 5.809 & 6.725 \\ \hline
 6.908 & 7.601 & \!\!-10.82 & 6.908 & \!\!-10.82 & 6.215 \\ \hline
\end{tabular}
}\\

\noindent
Using the log-transform, the multiplication operations in our deviation payoff calculation become additions and the exponents become multiplications.
As a first step, we need to take the natural log of our mixture probabilities.

\[ \ln \bigdot \left( \threevec{0.1}{0.5}{0.4} \right) = \threevec{-2.3026}{-0.6931}{-0.9163} \]

\noindent
Then we column-wise multiply the log probabilities by the configurations.\\

\noindent
\begin{tabular}{|C{10mm}|}
\hline
\!\!-2.3026 \\ \hline
\!\!-0.6931 \\ \hline
\!\!-0.9163 \\ \hline
\end{tabular}
$\bigdot\times$
\begin{tabular}{|C{1mm}|C{1mm}|C{1mm}|C{1mm}|C{1mm}|C{1mm}|}
\hline
2 & 1 & 1 & 0 & 0 & 0 \\ \hline
0 & 1 & 0 & 2 & 1 & 0 \\ \hline
0 & 0 & 1 & 0 & 1 & 2 \\ \hline
\end{tabular}
$\;\;=\;\;$
\begin{tabular}{|C{8mm}|C{8mm}|C{8mm}|C{8mm}|C{8mm}|C{8mm}|}
\hline
\!\!-4.605 & \!\!-2.303 & \!\!-2.303 & 0 & 0 & 0 \\ \hline
0 &  \!\!-0.693 & 0 & \!\!-1.386 &  \!\!-0.693 & 0 \\ \hline
0 & 0 & \!\!-0.916 & 0 & \!\!-0.916 & \!\!-1.833 \\ \hline
\end{tabular}\\

\noindent
Next we sum down each column and add the weighted payoffs.\\

\noindent
\begin{tabular}{|C{9.5mm}|C{9.5mm}|C{9.5mm}|C{9.5mm}|C{9.5mm}|C{9.5mm}|}
\hline
-4.605 & -2.996 & -3.219 & -1.386 & -1.609 & -1.833 \\ \hline
\multicolumn{6}{c}{$.+$} \\ \hline
6.215 & 6.502 & 6.502 & 6.502 & 7.195 & 6.502 \\ \hline
6.725 & 5.809 & 7.419 & 6.215 & 5.809 & 6.725 \\ \hline
6.908 & 7.601 & -10.82 & 6.908 & -10.82 & 6.215 \\ \hline
\multicolumn{6}{c}{$=$} \\ \hline
1.609 & 3.507 & 3.283 & 5.116 & 5.586 & 4.670 \\ \hline
2.120 & 2.813 & 4.200 & 4.828 & 4.200 & 4.893 \\ \hline
2.303 & 4.605 & -14.04 & 5.521 & -12.43 & 4.382 \\ \hline
\end{tabular}\\

\noindent
Now we can exponentiate to undo the log transform and then sum over the rows.\\

\noindent
\begin{minipage}{.1\columnwidth}
\vfill
\[ e\bigdot^{\wedge} \left( \; \boldsymbol{\cdot} \; \right) \]
\vfill
\end{minipage}
\begin{minipage}{.5\columnwidth}
$\;=$
\begin{tabular}{|C{9mm}|C{9mm}|C{9mm}|C{9mm}|C{9mm}|C{9mm}|}
\hline
\!5.0000 & \!33.333 & \!26.667 & \!166.67 & \!266.67 & \!106.67 \\ \hline
\!8.3333 & \!16.667 & \!66.667 & \!125.00 & \!66.667 & \!133.33 \\ \hline
\!10.000 & \!100.00 & \!8.0e-7 & \!250.00 & \!4.0e-6 & \!80.000 \\ \hline
\end{tabular}
\end{minipage}
\begin{minipage}{.3\columnwidth}
\[ \sum_{rows} \left( \; \boldsymbol{\cdot} \; \right) = \threevec{ 605.0000039}{416.6666725}{440.0000056} \]
\end{minipage}\\

\noindent
This leaves us with deviation payoffs for the re-scaled game from $10^{-5}$ to $10^3$, which are fine (in fact preferable) for computing equilibria, but if we want payoffs from the original game, we can undo the affine transformation, resulting in the usual answer of $\left[ 0.63 \; -\!0.50 \; -\!0.36 \right]^\mathrm{T}$.

\subsection{Further Improvements}

Subsequently, we proposed GPU acceleration, which performs a mathematically identical calculation to that shown for log-probabilities above, but takes advantage of SIMD instructions.
We also proposed operating on a batch of mixed strategies stored as a 2-dimensional array.
This is harder to show in a paper-and-pencil example, because it adds a third dimension to several of the arrays, and really amounts only to doing several instances of the same computation using instructions broadcast along that extra dimension.
The Julia code appendix shows how the implementations differ.

\vfill

\pagebreak

\section{Appendix B: Julia Code}

This appendix shows a small subset of the code implementing our data structures and Nash computation algorithms,
focusing on those parts that best illustrate the key concepts discussed in the paper.
For more complete (and runnable) implementations, see the code supplement.

\subsection{Data Structure Implementations}

First, we show the \texttt{deviation\_payoffs} function for the data structure variant from the Opponent Configurations section.
This is not the best version of the data structure, but it most-clearly exhibits the core idea of the vectorized deviation-payoff calculation.
The \texttt{@reduce} macro comes from the \texttt{TensorCast.jl} package.

{\small
\begin{verbatim}
function deviation_payoffs(game::SymGame_OppConfig, mixture::AbstractVector)
    @reduce probs[c] := prod(a) mixture[a] ^ game.config_table[a,c]
    @reduce dev_pays[a] := sum(c) game.payoff_table[a,c] * probs[c] * game.repeat_table[c]
    return dev_pays
end
\end{verbatim}
}

\noindent
Next we show the \texttt{struct} definition for the best version of the data structure, called \texttt{SymGame\_GPU}.
The \texttt{offset} and \texttt{scale} attributes are saved so that if necessary we can undo the affine transformation used to normalize the payoffs and present payoffs on their original scale.
Of note, all of the code for the \texttt{SymGame\_CPU} variant is identical, except that it uses \texttt{Array}s in place of \texttt{CUDA.CuArray}s.

{\small
\begin{verbatim}
struct SymGame_GPU <: SymmetricGame
    num_players::Int32
    num_actions::Int32
    config_table::CUDA.CuArray{Float32,2}
    payoff_table::CUDA.CuArray{Float32,2}
    offset::Float32
    scale::Float32
end
\end{verbatim}
}

\noindent
The constructor for the \texttt{SymGame\_GPU} struct takes a \texttt{payoff\_generator} function as input.
This is a function that, when passed a configuration $\config$, returns a vector of payoffs $\vec{\purepay}$ containing $\purepay_a(\config)$ for each action $a$.
\texttt{CwR} is an alias for \texttt{Combinatorics.with\_replacement\_combinations}.
The call to \texttt{logmultinomial} uses \texttt{SpecialFunctions.loggamma} to compute the log of the multinomial from Equation~\eqref{eq:dev_reps}.
The call to \texttt{set\_scale} determines the positive affine transformation that will normalize the payoffs to the range \texttt{const MIN\_PAYOFF = 1e-5} through \texttt{const MAX\_PAYOFF = 1e3}.

{\small
\begin{verbatim}
function SymGame_GPU(num_players, num_actions, payoff_generator)
    num_configs = multinomial(num_players-1, num_actions-1)
    config_table = Array{Float32}(undef, num_actions, num_configs)
    payoff_table = Array{Float32}(undef, num_actions, num_configs)
    log_repeat_table = Array{Float32}(undef, 1, num_configs)
    for (c,config) in enumerate(CwR(1:num_actions, num_players-1))
        prof = counts(config, 1:num_actions)
        config_table[:,c] = prof
        log_repeat_table[c] = logmultinomial(prof...)
        payoff_table[:,c] = payoff_generator(prof)
    end
    (offset, scale) = set_scale(minimum(payoff_table), maximum(payoff_table))
    payoff_table = log.(normalize(payoff_table, offset, scale)) .+ log_repeat_table
    SymGame_GPU(num_players, num_actions, config_table, payoff_table, offset, scale)
end
\end{verbatim}
}

\noindent
Next, we show the function that computes deviation payoffs on a batch of mixtures.
This differs from the \texttt{OppConfig} version above in that most operations are performed under a log transformation, and that the computation is being broadcast across an additional mixture-batch dimension.
Note that the returned deviation payoffs are in the normalized scale, because this helps with hyperparameter tuning for various Nash algorithms.
A call to \texttt{denormalize(game, payoffs)} can transform them back to the game's original payoff scale.

{\small
\begin{verbatim}
function deviation_payoffs(game::SymGame_GPU, mixture_batch::CUDA.CuArray{Float32,2})
    log_mixtures = log.(mixture_batch .+ eps32)
    @reduce log_probs[m,c] := sum(a) log_mixtures[a,m] * game.config_table[a,c]
    @reduce dev_pays[a,m] := sum(c) exp(game.payoff_table[a,c] + log_probs[m,c])
    return Array(dev_pays)
end
\end{verbatim}
}

\noindent
Finally, we show the function to compute deviation derivatives for a single mixture.
Note that the \texttt{mixture .+ eps32} operation ends up only changing array entries equal to zero, and is slightly more efficient than explicitly filtering for those entries.

{\small
\begin{verbatim}
function deviation_derivatives(game::SymGame_GPU, mixture::CUDA.CuArray{Float32,1})
    mixture = mixture .+ eps32
    log_mixture = log.(mixture)
    @reduce log_probs[c] := sum(a) log_mixture[a] * game.config_table[a,c]
    deriv_configs = game.config_table ./ mixture
    @reduce J[a,s] := sum(c) exp(game.payoff_table[a,c] + log_probs[c]) * deriv_configs[s,c]
    return J
end
\end{verbatim}
}

\noindent
The code supplement implements many additional features for this and other data structure variants, including batched deviation derivatives, single-mixture deviation payoffs, and a number of other convenience functions.

\subsection{Nash Algorithm  Implementations}

Here, we present code for the two algorithms we have found to be the most practical for identifying Nash equilibria in large symmetric games: replicator dynamics and gradient descent on sum-of-gains.
Both of these functions can operate on either a vector representing a single mixture or an array representing a batch of mixtures, thanks to multiple dispatch to vector- or matrix-versions of the \texttt{deviation\_payoffs} and \texttt{gain\_gradients} functions, respectively.

For replicator dynamics, the \texttt{offset} input allows for the possibility that the game's payoffs are not normalized, in which case it should be less than or equal to the minimum payoff in the game to prevent negative probabilities.
This parameter can also be used as a quasi-step-size, since a larger negative offset will diminish the relative magnitude of payoff differences.
The \texttt{simplex\_normalize} function simply divides by the sum to ensure probabilities add to 1.

{\small
\begin{verbatim}
function replicator_dynamics(game::SymmetricGame, mix::AbstractVecOrMat;
                             iters::Integer=1000, offset::Real=0)
    for i in 1:iters
        mix = simplex_normalize(mix .* (deviation_payoffs(game, mix) .- offset))
    end
    return mix
end
\end{verbatim}
}

\noindent
For gain descent, we allow either a fixed step size or schedule of step sizes.
The \texttt{simplex\_normalize} function implements the projection algorithm given by \citet{simplex_projection}, and also operates on either a single mixture or a batch.

{\small
\begin{verbatim}
function gain_descent(game::SymmetricGame, mix::AbstractVecOrMat;
                      iters::Integer=1000, step_size::Union{Real,AbstractVector}=1e-6)
    if step_size isa Real
        step_size = [step_size for i in 1:iters]
    end
    for i in 1:iters
        mix = simplex_project(mix .- step_size[i]*gain_gradients(game, mix))
    end
    return mix
end
\end{verbatim}
}

\noindent
Here is the implementation of the \texttt{gain\_gradients} function for a single mixture.
The version operating on a batch of mixtures also appears in the code supplement.

{\small
\begin{verbatim}
function gain_gradients(game::SymmetricGame, mixture::AbstractVector)
    dev_pays = deviation_payoffs(game, mixture)
    mixture_EV = mixture' * dev_pays
    dev_jac = deviation_derivatives(game, mixture)
    util_grads = (mixture' * dev_jac)' .+ dev_pays
    gain_jac = dev_jac .- util_grads'
    gain_jac[dev_pays .< mixture_EV,:] .= 0
    return dropdims(sum(gain_jac, dims=1), dims=1)
end
\end{verbatim}
}

\noindent
The code supplement also contains implementations of fictitious play and iterated better-response, along with a number other functions including regret-filtering and deduplication for processing candidate equilibria returned by these methods.

\vfill

\pagebreak

\section{Appendix C: Nash Algorithms}

In this appendix we provide a description of a few additional Nash-finding algorithms, formulated in terms of deviation payoffs.
Note that this is by no means a complete survey of Nash-finding algorithms.

\paragraph{Fictitious play} \cite{fictitious_play} operates by iteratively best-responding to the opponents' empirical distribution of past actions as though that represented a mixed strategy.
In the context of a symmetric game, we can formulate FP as keeping a count of the number of times each action has been ``played'', and normalizing those counts to produce a mixed strategy.
Then on each iteration, the count for the best response to the current mixture is ``played'', incrementing that action's count.
Because we need to specify some initial distribution for the first step, we begin with a mixture $\mix^0$, and generate the initial counts based on some constant multiple of those probabilities: $\vec{w} \gets c \cdot \mix^0$.

\begin{align*}
b &\gets \max_{a} \devpay_a(\mix^t) \\
\vec{w}_b &\gets \vec{w}_b + 1 \\
\mix^{t+1} &\gets \frac{1}{\sum_a \vec{w}_a} \vec{w}
\end{align*}

\paragraph{Scarf's simplicial subdivision} algorithm \cite{Scarf} is a generic method for finding the fixed point of a continuous function on the simplex.
Since the existence of Nash equilibria is proven by constructing a continuous function whose fixed-points correspond to equilibria, we can apply Scarf's algorithm to that function.
Similar to the sum-of-gains $g$ we defined for gradient descent, we can define the gain for a particular action, and use it to construct the better-response function $B$.

\begin{align*}
g_a(\mix) &= \max \left( 0 , \devpay_a(\mix) - \devpay(\mix) \cdot \mix \right)  \\
B_a(\mix) &= \frac{\mix_a + g_a(\mix)}{1 + \sum_a g_a(\mix)}
\end{align*}

The function $B$ outputs a mixture that increases the relative probability of any actions that have deviation payoffs better than the expected utility of $\mix$.
Nash equilibria of the game correspond exactly to fixed-points of $B$, so we can use Scarf's algorithm to identify a fixed point and therefore a Nash equilibrium.
Scarf's algorithm traces a path through sub-spaces of the simplex to identify one that must contain a fixed point.
Then within the identified sub-simplex, further subdivisions can be searched recursively to narrow in on a fixed point.

An even simpler algorithm that makes use of $B$ is to simply apply it repeatedly.
We refer to this algorithm as iterated better-response.
This algorithm tends to behave like a smoothed version of fictitious play, because it up-weights several better responses instead of just a single best response.
Figure~\ref{fig:fp_ibr} in the Appendix D shows traces of FP and IBR on the same game.
Note that these algorithms follow very similar trajectories, except that FP always moves in a straight line toward one of the corners of the simplex, since it only up-weights one action, while IBR can curve, by up-weighting multiple actions at once, and by slightly different amounts on consecutive iterations.

For all of the incomplete algorithms described (replicator dynamics, gradient descent, fictitious play, and iterated better-response), we typically run them for a fixed number of iterations, from a grid of equally-spaced starting mixtures.
The resulting mixtures may not be Nash equilibria, due to non-convergence or local-optimality.
We therefore filter out any endpoints with regret above a predetermined $\varepsilon$.
It is also common for several runs to converge to essentially the same $\varepsilon$-equilibrium, so we filter the results to eliminate ``duplicate'' points within some threshold for Euclidean or Chebyshev distance.

\pagebreak

\section{Appendix D: Supplementary Figures}

In this appendix we provide, for completeness, a number of slight variations on the figures appearing in the main text.

\begin{figure}[h]
\center
\includegraphics[width=.5\textwidth]{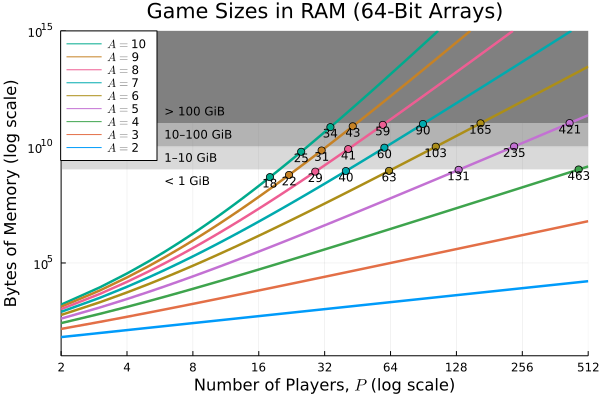}
\caption{RAM required to store the CPU version of our data structure. Compared to Figure~\ref{fig:mem_limit}, this shows sizes using arrays of 64-bit floats. Note that all of the points indicating the largest instance below a given memory threshold have just slightly fewer players than the corresponding 32-bit variant, because 64-bit arrays double the memory usage, while the arrays grow combinatorially in players/actions.}
\end{figure}

\begin{figure}[h]
\center
\includegraphics[width=.49\textwidth]{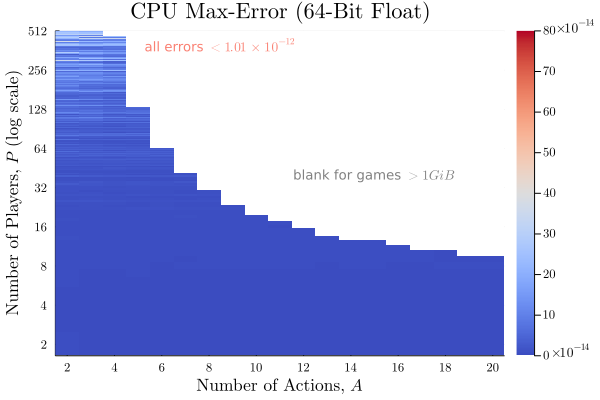}
\hfill
\includegraphics[width=.49\textwidth]{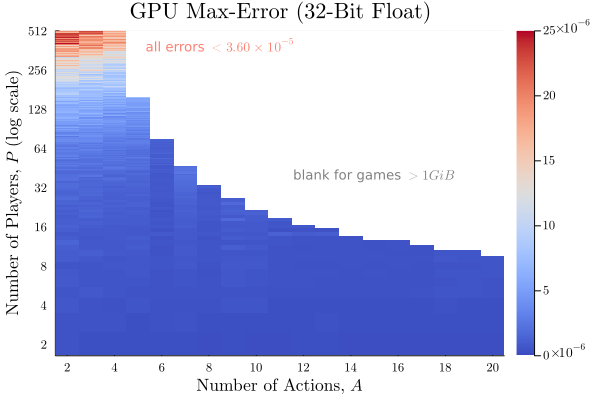}
\caption{Heatmaps showing the largest deviation-payoff errors encountered across 1000 mixtures in 10 random games at each combination of $P$ and $A$. We consider nearly all of these errors to be entirely negligible, with the possible exception of $P > 256$ for 32-bit arrays. This suggests that for extremely large numbers of players, the added precision of 64-bit arithmetic may be worthwhile.}
\label{fig:errors}
\end{figure}

\begin{figure}[h]
\center
\includegraphics[width=.49\textwidth]{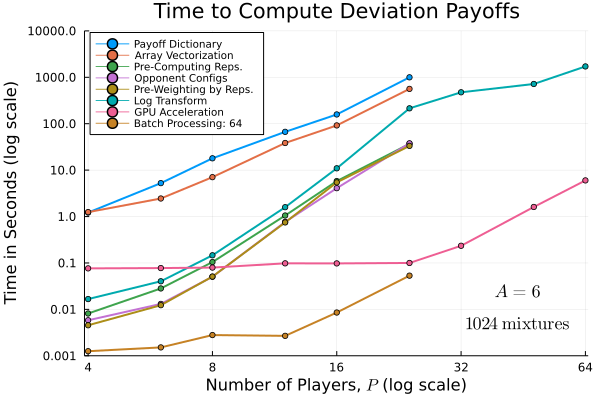}
\hfill
\includegraphics[width=.49\textwidth]{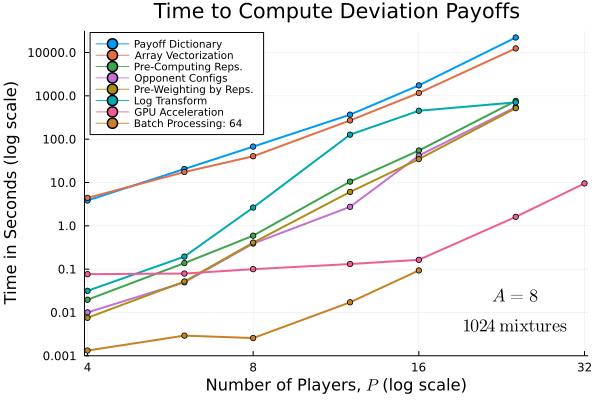}
\caption{Time to required to compute deviation payoffs for 1024 mixtures in 6-action (left) and 8-action (right) games using each data structure variant. Lines stop when either an integer overflow is encountered or when more than 1GB of memory is required. The test for $A=8$ and $P=64$ with the log-transform data structure timed out. Note that both x- and y-axes have been rescaled relative to Figure~\ref{fig:dev_pay_timing}.}
\end{figure}

\begin{figure}[h]
\center
\includegraphics[width=.49\textwidth]{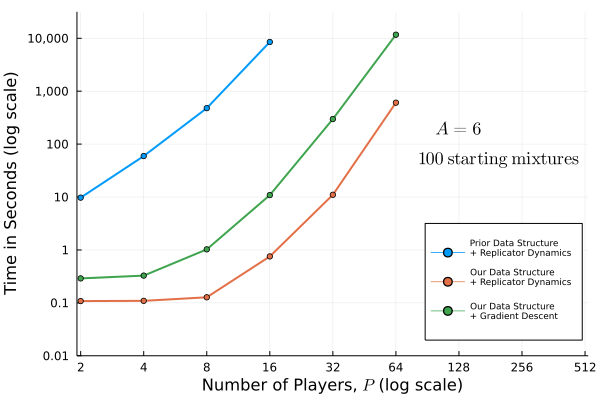}
\hfill
\includegraphics[width=.49\textwidth]{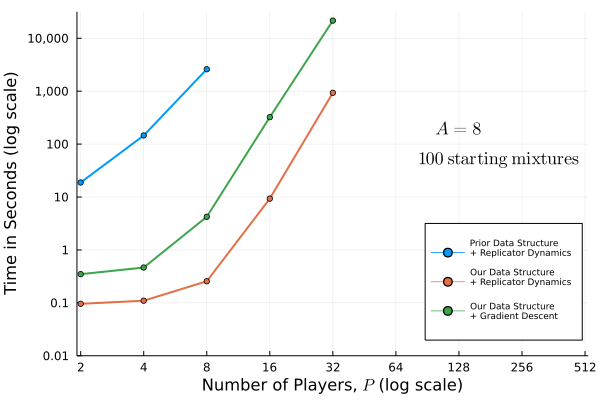}
\caption{Time to compute Nash equilibria in 6-action (left) and 8-action (right) games with 100 starting mixtures and 1000 iterations. Note that relative to Figure~\ref{fig:eq_timing} we have much slower computations for any given number of players, and the lines all terminate at smaller numbers of players due to reaching the 1GiB memory limit we enforced in these experiments.}
\end{figure}

\begin{figure}[h]
\center
\includegraphics[width=.49\textwidth]{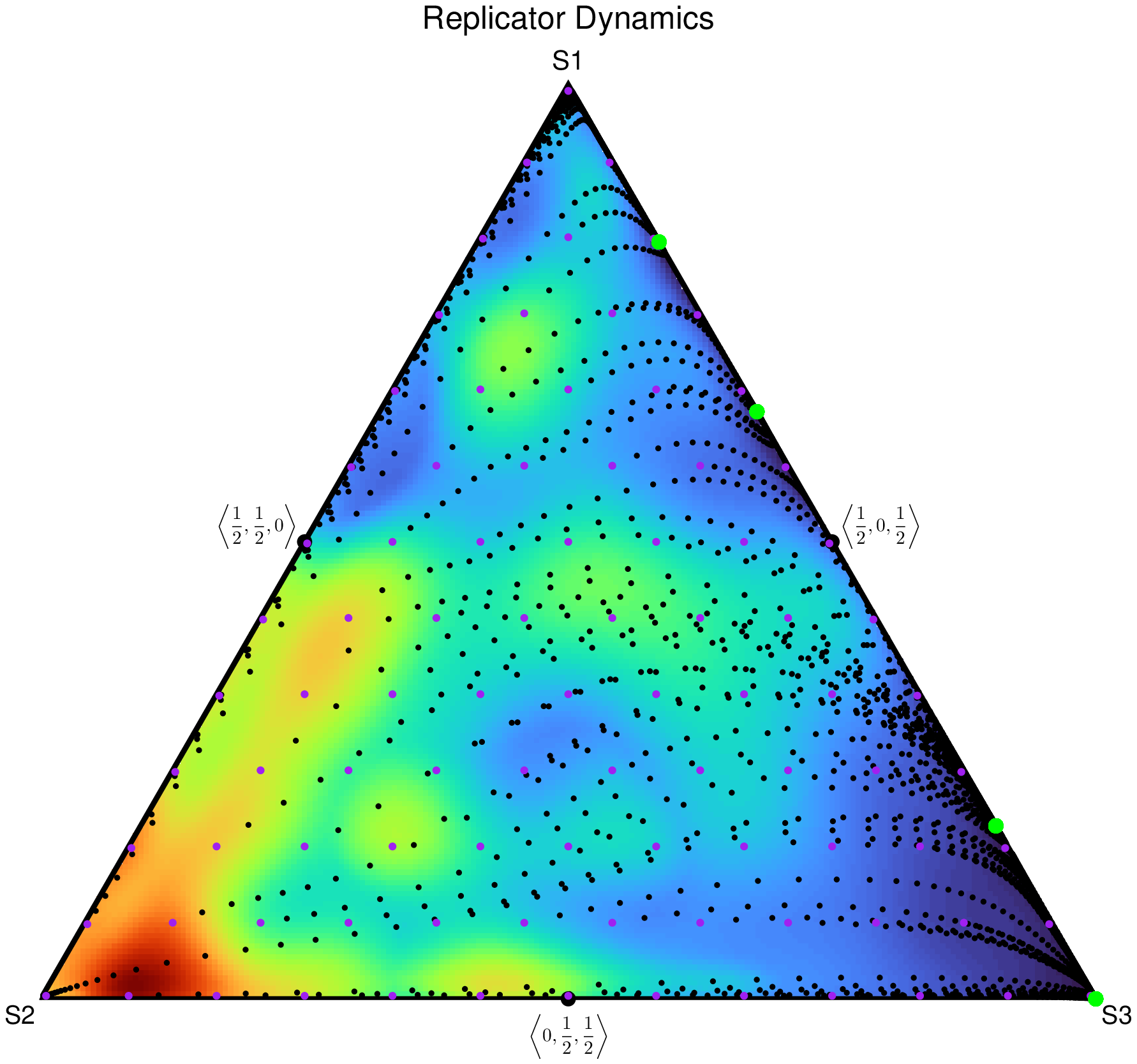}
\hfill
\includegraphics[width=.49\textwidth]{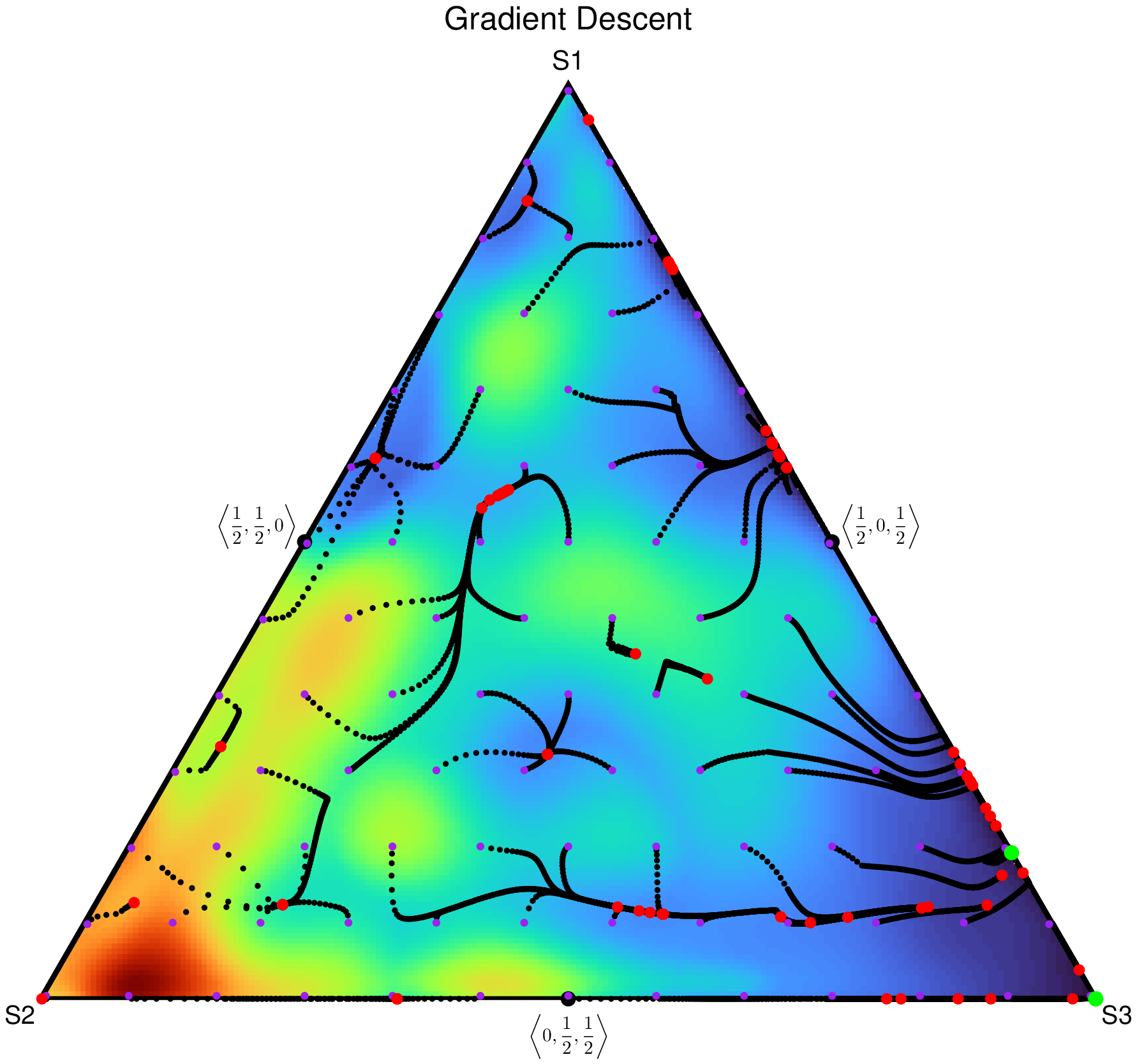}
\caption{A game where replicator dynamics (left: green points indicating 4 $\varepsilon$-Nash equilibria identified) outperforms gradient descent (right: green points indicating 2 $\varepsilon$-Nash identified). Note that a number of GD runs ended up close to the other two equilibria found by RD, so with better-tuned hyperparameters, they may have been identified.}
\label{fig:rd_optimal}
\end{figure}

\begin{figure}[h]
\center
\includegraphics[width=.49\textwidth]{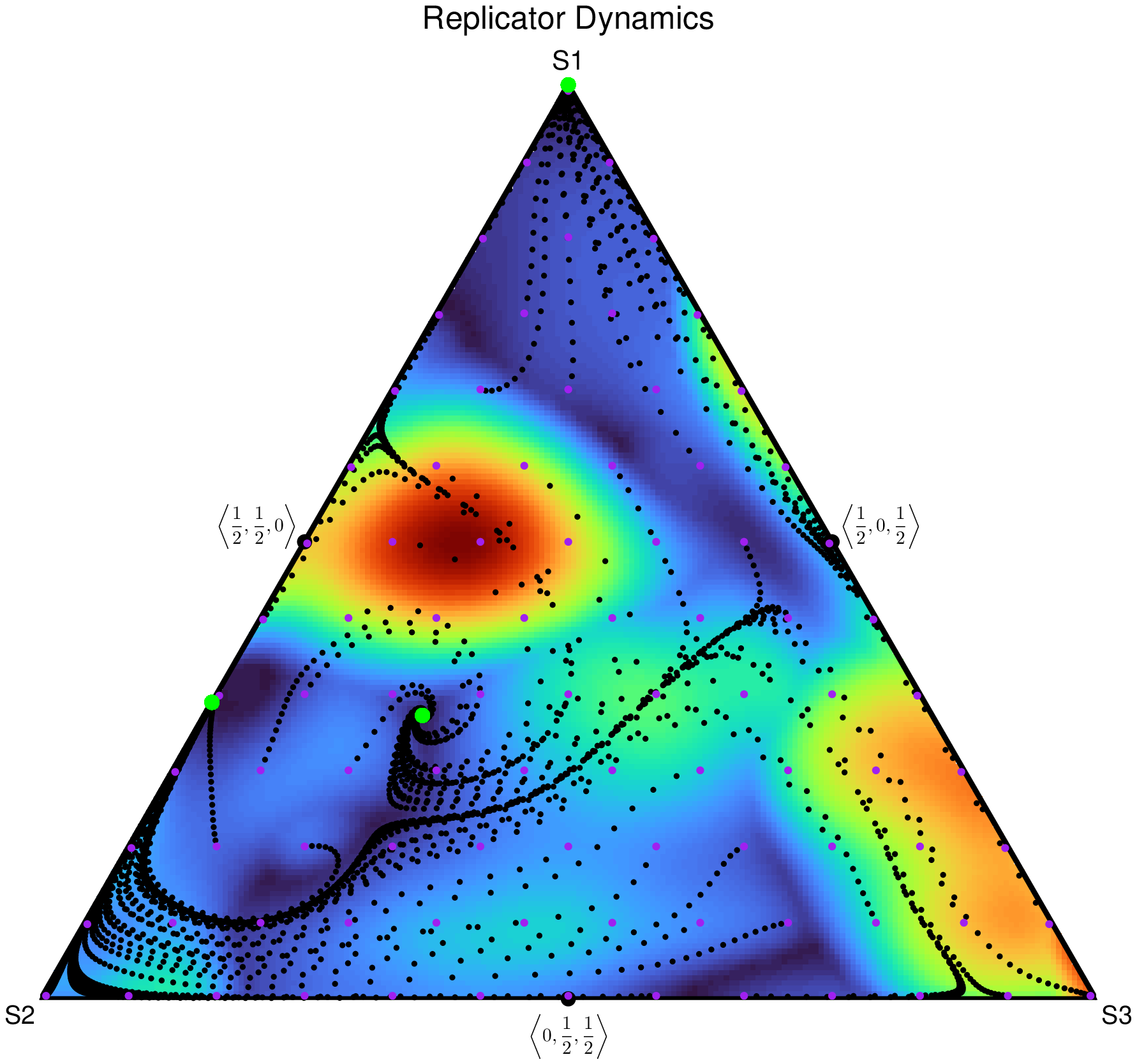}
\hfill
\includegraphics[width=.49\textwidth]{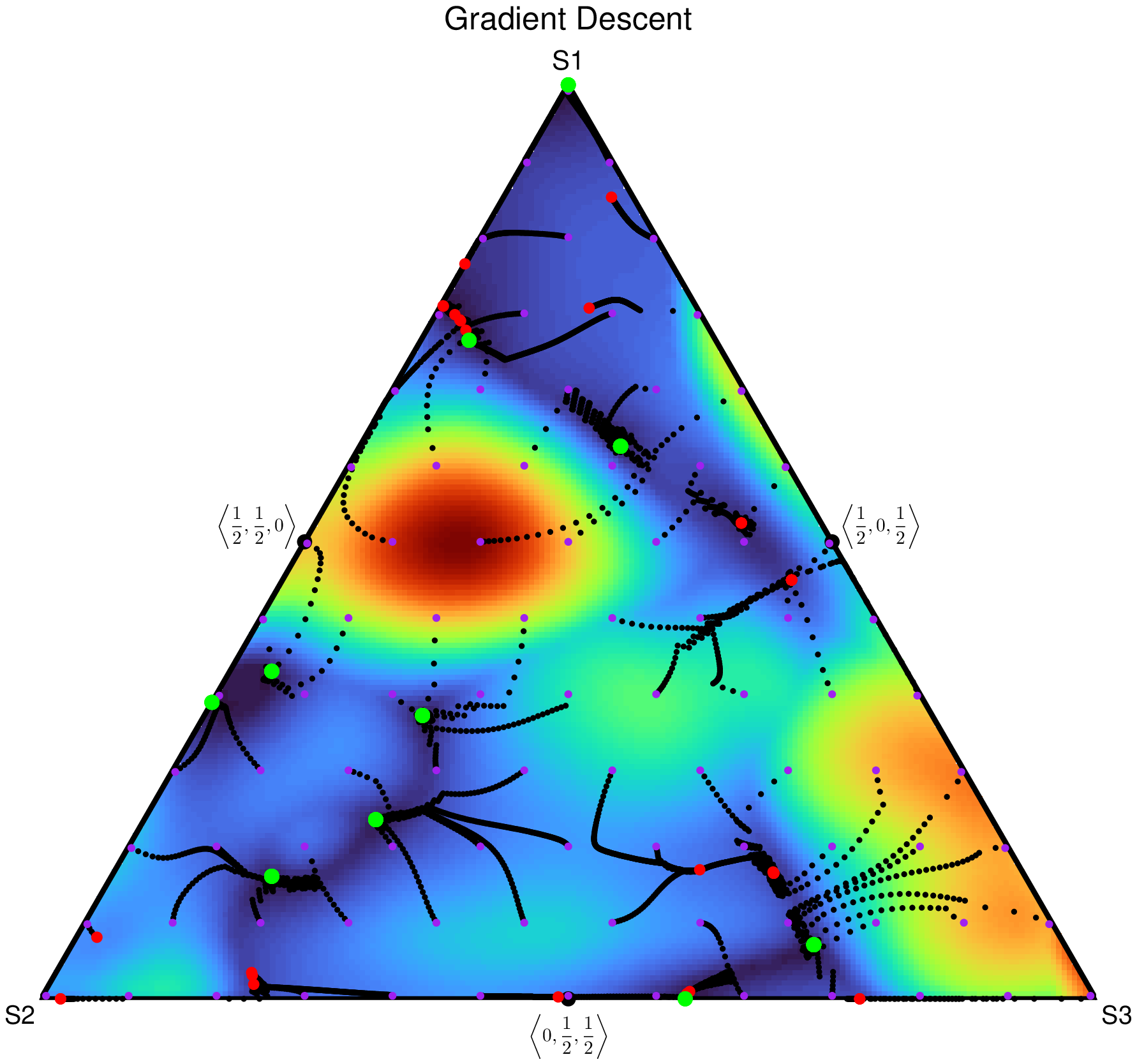}
\caption{A game where gradient descent (right: green points indicating 10 $\varepsilon$-Nash identified) outperforms replicator dynamics (left: green points indicating 3 $\varepsilon$-Nash equilibria identified). Note that GD tends to have many more failed runs (ending at red points), but that RD tends to have much more lopsided basins of attraction, and can therefore miss certain equilibria entirely.}
\label{fig:gd_optimal}
\end{figure}

\begin{figure}[h]
\center
\includegraphics[width=.49\textwidth]{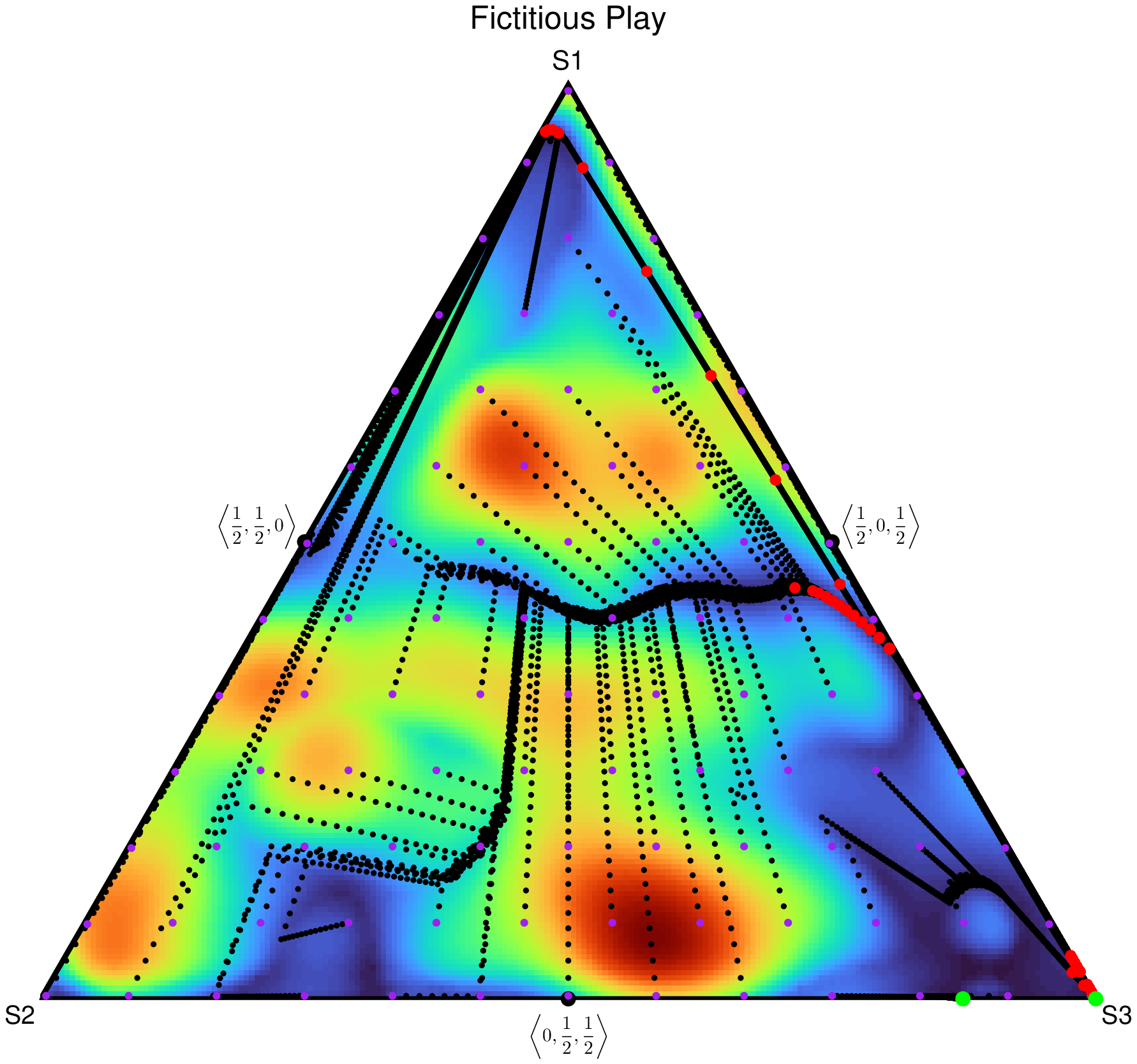}
\hfill
\includegraphics[width=.49\textwidth]{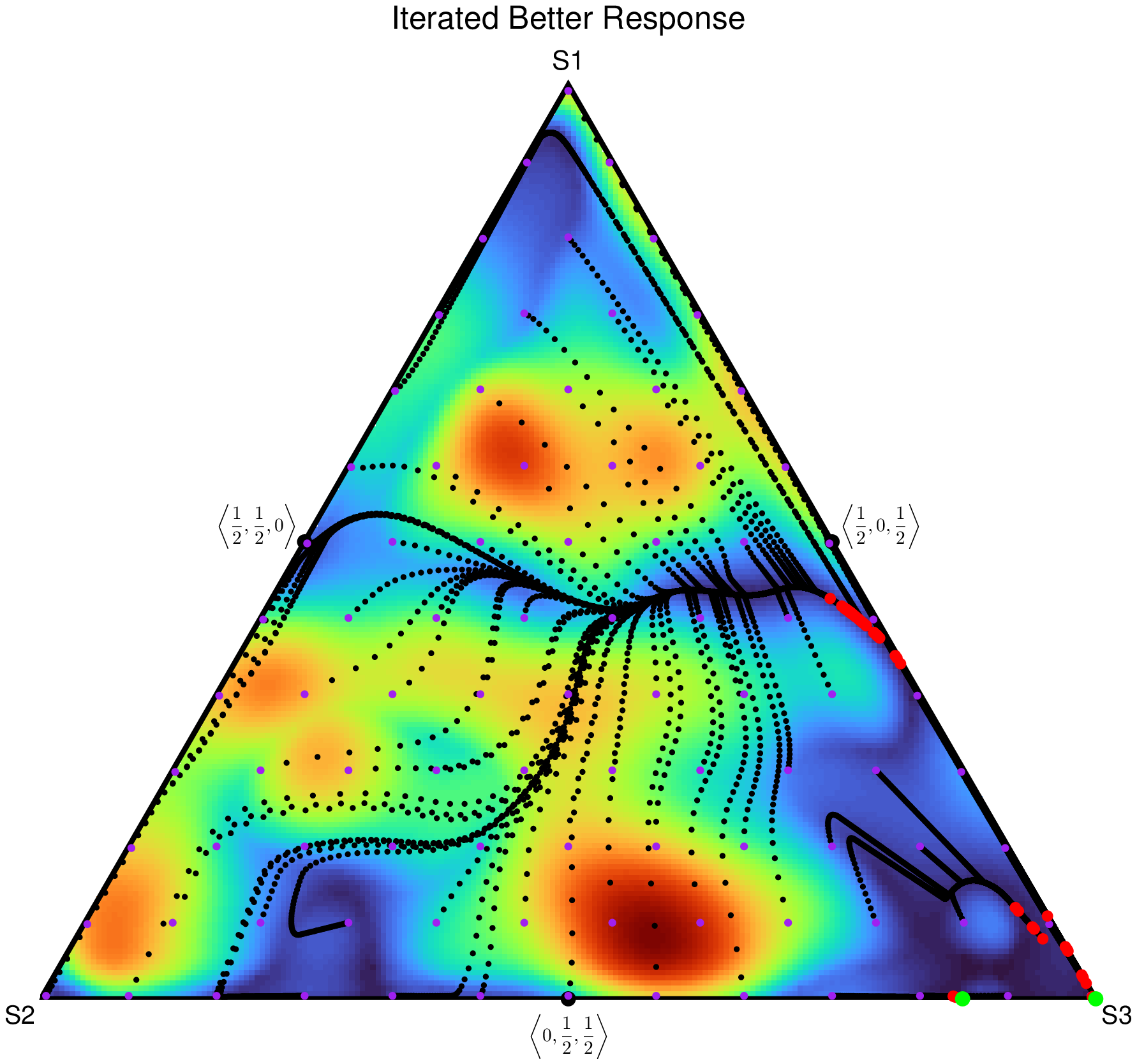}
\caption{Typical performance of fictitious play and iterated better-response. Both algorithms tend to struggle with converging to mixed-strategy equilibria.}
\label{fig:fp_ibr}
\end{figure}

\begin{figure}[h]
\center
\includegraphics[width=.5\textwidth]{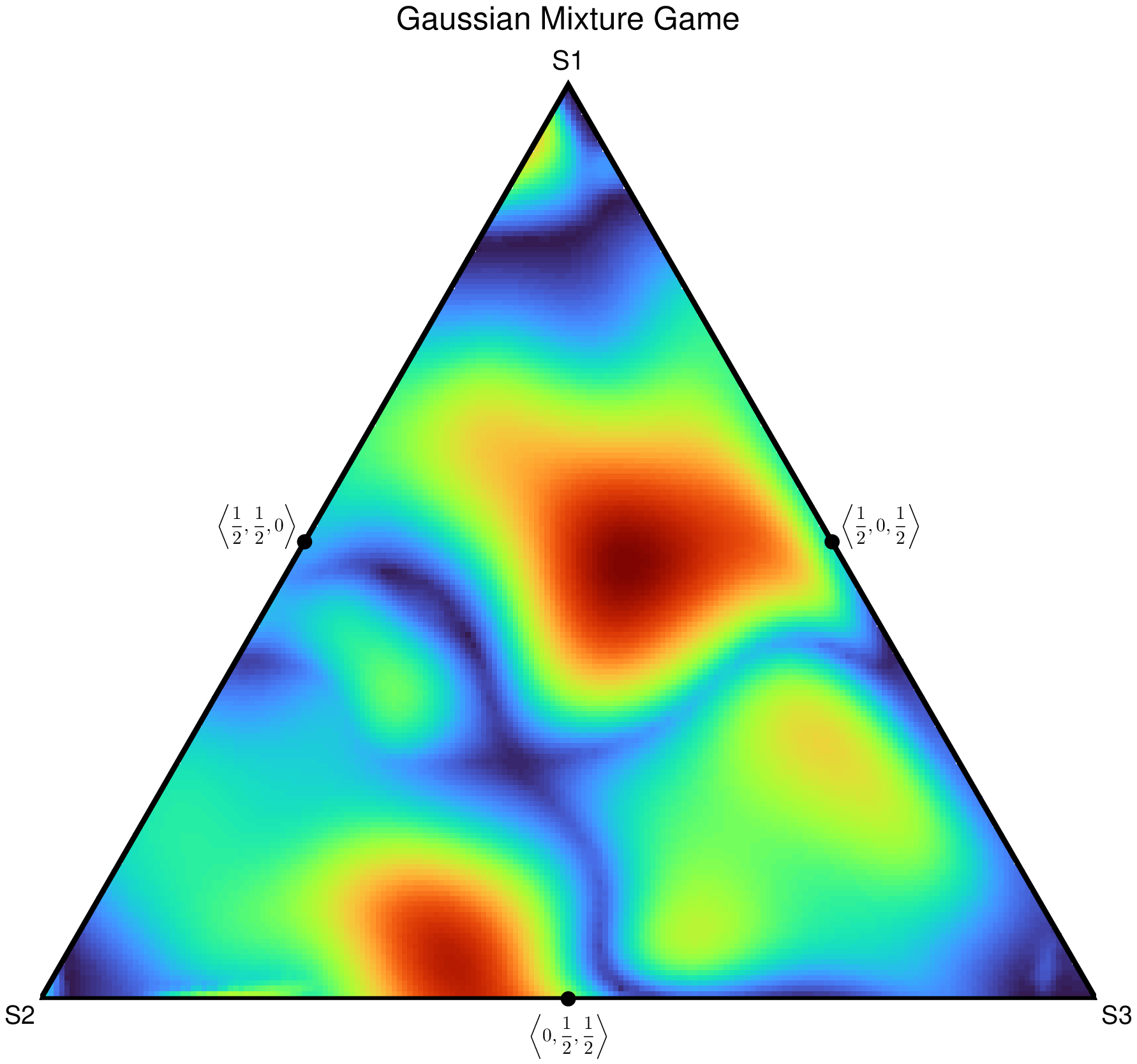}
\caption{A regret heatmap for a randomly generated 3-action Gaussian mixture game.}
\label{fig:gmg}
\end{figure}

\pagebreak

Most of the simplex plots shown here are based on random additive-sine games.
We generate these games as bipartite action-graph games with 200 function nodes.
Each action will be connected to a random subset of the function nodes by an Erd\H{o}s-Renyi random-graph process.
Then each function node is connected to every action with a random weight.
Function nodes are contribution-independent and simply count the number of agents in their neighborhood.
They then output a value based on a random sin-function added to a random low-degree polynomial.
Each action then computes its value as a weighted sum over the function outputs.

However, our data structures make it easy to specify even more interesting classes of random games.
As an example, we generated random Gaussian mixture games for some of our experiments.
To generate a GMG, we specify a number of functions per action, and for each, we choose a random centroid from a Dirichlet distribution, scaled to the size-$P$ integer simplex, a random covariance matrix from a Lewandowski-Kurowicka-Joe distribution, and a random multiplicative scaling factor (since these Gaussians need not represent valid probability distributions).
Then to determine the payoffs for a given configuration/action pair, we evaluate each Gaussian function for that action at the given configuration, multiplied by the scaling factor, and sum the results.
An example regret heatmap for a 3-action Gaussian mixture game is shown in Figure~\ref{fig:gmg}.

\end{document}